\documentclass[a4paper,aps,pre,superscriptaddress,floatfix,nofootinbib,twocolumn,noshowpacs,bibtex]{revtex4}

\usepackage[pdftex]{graphicx}
\usepackage{latexsym,amsmath,verbatim}
\usepackage{color}
\usepackage{rotating}
\usepackage{multirow}
\usepackage[english]{babel}

\begin{document}

\title{In science ``there is no bad publicity'': Papers criticized in comments have high scientific impact}

\author{Filippo Radicchi}\email{f.radicchi@gmail.com}
\affiliation{Departament d'Enginyeria Quimica, Universitat Rovira i Virgili, Av. Paisos Catalans 26, 43007 Tarragona, Catalunya, Spain}

\begin{abstract}
\noindent Comments are special types of publications 
whose aim is to correct or criticize previously published papers. For this reason, comments are believed to make commented papers less worthy or trusty to the eyes of the scientific community, and thus predestined to have low scientific impact. Here, we show that such belief is not supported by empirical evidence. We consider thirteen major publication outlets in science, and perform systematic comparisons between the citations accumulated by commented and non commented articles. We find that (i) commented papers are, on average, much more cited than non commented papers, and (ii) commented papers 
are more likely to be among the most cited papers of a journal. 
Since comments are published soon after criticized papers, comments should be viewed as early indicators of the future impact of criticized papers.
\end{abstract}


\maketitle

\section{Introduction}
\noindent In $1543$, Nikolaus Copernicus proposed a heliocentric model
that will have revolutionized the human view of the universe~\cite{Kuhn1957}. 
The paradigm shift proposed by the new
ideas of Copernicus caused an unavoidable
controversy in the scientific community of his era, 
still anchored to the geocentrism. The dispute involved some of
the brightest brains of the period
--including Galilei, Kepler and Netwon-- and lasted for more than
two centuries before the complete
acceptance of the heliocentric
model for the the description of the solar system. 
The controversy behind the Copernican revolution is 
just one of the most popular 
examples of scientific controversies
that are part of the history of science.
Other well known examples are
the controversy which followed
the publication of the theory of evolution by Darwin~\cite{Vorzimmer1972}, the 
Bohr-Einstein debate about the fundamentals
of quantum mechanics~\cite{AlbertEinstein1949}, and the dispute
originated by Wegener with his theory of 
continental drift~\cite{Jacoby1981}. In the course 
of the history of science, however, not all scientific disputes 
have been resolved in favor of the original idea 
that caused the controversy:
the observation of $N$-rays~\cite{Klotz1980}, the theory 
of cold fusion~\cite{Lewenstein1992} and
the finding of water memory~\cite{Davenas1988,Hirst1993}
are all examples of theories or experimental results, associated to
fervent scientific controversies,
that have been at the end rejected
or disregarded by the scientific community.

\noindent Either resolving in favor
or against the scientific findings that
originated the disputes, scientific controversies are 
thought to be necessary for scientific 
progress~\cite{Engelhardt1987, Machamer2000}.
Even if not all the greatest achievements
in science have passed through a dispute, 
as for example the unification of electricity 
and magnetism by Maxwell, many major steps in 
science have been controversial.
Revolutionary changes
are {\it per se} controversial because they
reverse previous scientific paradigms, and thus
necessarily encounter some resistance 
before getting accepted. Scientific disputes, 
however, are not only associated to revolutionary 
discoveries, but they are also part of the 
process of scientific production: science
is, in fact, simultaneously 
a cooperative and  antagonistic enterprise,
where scientists both collaborate,
with the interchange of information, and compete,
through the exchange of criticisms. 
While the structure of collaboration
networks~\cite{Newman2004}
and the importance of teams 
for the creation of scientific 
knowledge~\cite{Guimera2005,Wuchty2007} have been 
empirically analyzed, no much
is quantitatively known about
scientific disputes. Scientific controversies are
usually studied in philosophy of science,
but only through the analysis of popular case
examples and never in quantitative 
terms~\cite{Engelhardt1987, Machamer2000}.    
\\
\noindent Here for the first time, we provide a quantitative and
large-scale study of scientific controversies. 
We focus our attention on modern scientific disputes identifiable with
the publication of formal comments. 
We systematically study the difference 
between the citations accumulated
by commented and non commented papers, and show
that comments
can be statistically interpreted
as early signs of the future impact of criticized
papers.



\section{Results}
\noindent Comments are short publications  whose purpose is to address
core arguments, theories or experimental
results of recently published research 
papers. The name of this type of publications
varies from journal to journal. For example, in {\it Nature} they are
called ``Brief Communications Arising'', in {\it Science}
 ``Technical Comments'', in {\it Physical Review Letters}
 ``Comments'', in 
{\it New England Journal of Medicine} ``Letters to the Editor'', etc. 
When submitted, comments are considered by journal
editors in a similar way as normal articles:
in order to be published, they need to satisfy strict requirements of
broad interest, and have to pass the scrutiny
of peer-review. 
The vast majority of comments represent formal criticisms to
the content of commented papers.
Irrespectively of
the journal of publication, editorial policies are, 
in this regard, very explicit:
just to cite an example, in {\it Physical Review Letters} ``a comment
corrects or criticizes a specific Letter'' and
``is not meant to be a vehicle for addenda'' 
({\tt forms.aps.org/author/comments-prl.pdf}).
In addition, a comment is generally followed by a so-called ``reply''
or ``response'', written by the same authors of the criticized paper,
with the purpose of defending their own paper
from the criticisms of the comment. Comments and replies are
published one after the other in the same issue of the journal,
and thus provide a fair way to give
birth to a scientific dispute, where both parts present 
and defend their own opinions.
\\

\begin{figure}[!ht]
\begin{center}
\includegraphics[width=0.45\textwidth]{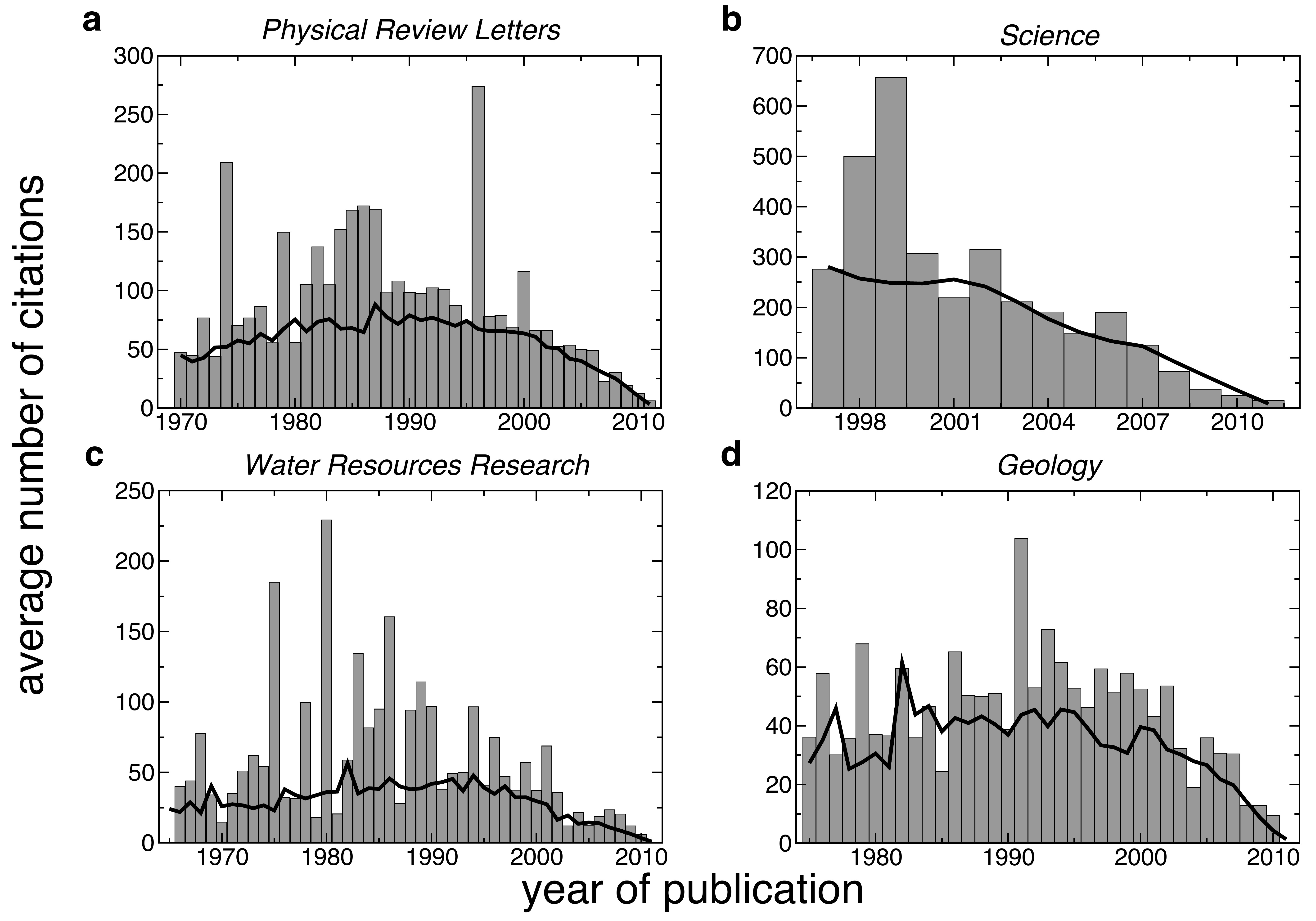}
\end{center}
\caption{
Commented papers accumulate on average
more citations. {\bf a.} Average number of citations received by papers published
in {\it Physical Review Letters} as a function of the year
of publication. The average number of citations accumulated
by non commented papers is represented by the black line,
while the average number of citations received by commented
papers is represented by the gray bars. 
{\bf b}, {\bf c} and {\bf d.} Same as in panel a but
for articles published in {\it Science}, {\it Water Resources Research} and
{\it Geology}, respectively.
}
\end{figure}
\noindent We identified all formal comments
published
in the last ten to fifty years
in thirteen major publication outlets in science,
including multidisciplinary and specialist journals
(focusing on research topics in environmental sciences, 
geology, medicine and physics).
We automatically associated each comment 
to the criticized  publication (see supplementary information). 
Finally, we collected  
from the Web Of Science database ({\tt isiknowledge.com})
the number of citations accumulated by each publication
(as of April $2012$). In our citation 
analysis, we restricted
the attention only to papers published before $2008$ in order
to rank papers on the
basis of stable citation distributions~\cite{Stringer2008}
and be confident that the vast majority
of comments to these papers have been already published 
(i.e., in more than the $95\%$ of the cases, comments
are published less than $4$ years after criticized papers, 
see Figures~3 and S2-10).

\noindent Though commented papers represent a small percentage of
the publications of a journal (Table~S1), already
with a qualitative analysis
it is possible to notice that many 
of the top cited articles of a journal
are papers that were criticized in 
formal comments
(Tables~S2-4).
In  {\it Physical Review Letters} for example, 
the most cited paper 
is ``Generalized Gradient Approximation Made Simple'',
{\it Phys. Rev. Lett.} {\bf 77}, 3865 (1996), with over $20,000$ citations,
and this paper was criticized in a formal
comment. More in particular, 
while the percentage of commented papers in
 {\it Physical Review Letters} in our period of observation 
is just $3\%$, 
we find that the $5$ most cited commented papers
are in the list of the $16$ most cited papers of the journal, a
proportion $9.4$ times larger than what expected by chance.
In the other journals, the situation is similar 
and the $5$ most cited commented papers 
have an absolute rank, based on the raw number of citations
they have accumulated, from $2$ to $44$ times higher than what 
expected in the case in which being or not being commented would be
unrelated to 
the number of citations accumulated (Table S1).
The only 
less evident case  is represented by the publications
in {\it Nature},
where the rank of the $5$ most cited commented papers
of the journal is $1.1$ times higher than 
what expected by chance.

\noindent From the previous qualitative analysis
based on a limited number of case examples, it seems 
not only that
commented papers receive more citations than
non commented papers, but also that
commented articles are unexpectedly more present
in the population of the most cited publications. 
In order to statistically confirm the validity of these observations,
we present here a systematic analysis. 
\\
First, we look at the average citation
rates of commented and non commented papers. 
In order
to avoid age-dependent biases in the 
number of citations~\cite{Radicchi2008}, we compare the
average number of citations
received by papers published in the same year. 
The results for the various journals are presented in 
Figures~1 and~S2-10. We confirm indeed that
commented papers are, on average, much more cited than 
non commented papers. 
However, the observation that a commented paper
typically receives more citations than a 
non commented article
is valid not only by looking at
average citation numbers, but 
also by considering 
other types of measures (e.g, median citation numbers, Figure~S11)
and performing
non parametric statistical tests 
(e.g., Mann-Whitney-test, Figure~S12).
In general, 
while in single years of publication the fact
that a commented paper is more cited than a non commented
article
could be still explained in terms
of statistical fluctuations,
the persistence over many years of a positive
signal allows us to say that
commented papers significantly accumulate more citations
than non commented articles.
\\
\begin{figure}[!ht]
\begin{center}
\includegraphics[width=0.45\textwidth]{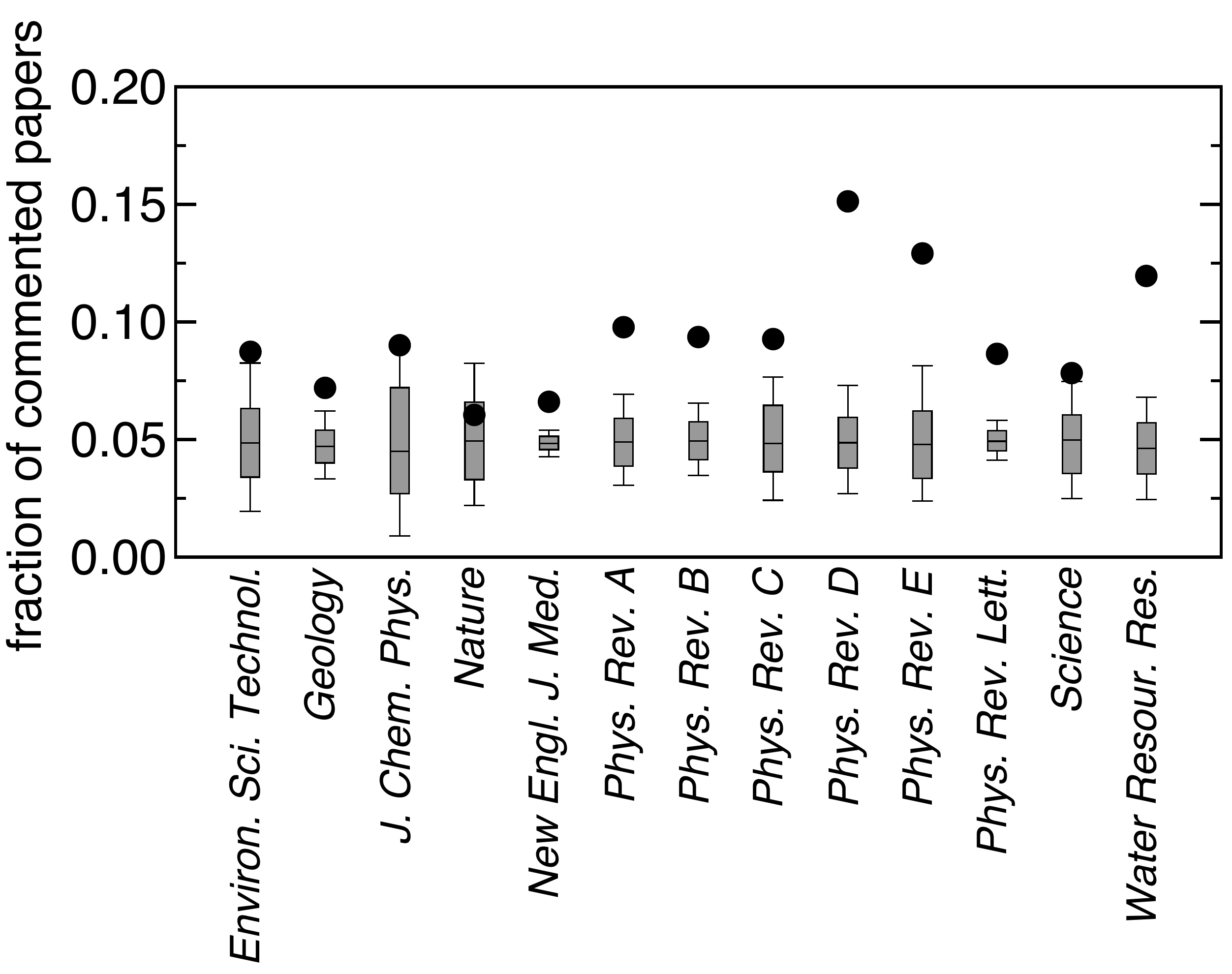}
\end{center}
\caption{
Commented papers are more likely to be 
among the most cited publications of a journal.
Fraction of commented papers (black circles) that are part
of the top $5\%$ of the most cited papers 
published in the journals
considered in our analysis.
This fraction is calculated as $C_{5\%} / C$, with 
$C_{5\%}$ number of commented papers within 
the top $5\%$, and $C$ total number of commented papers.
The quantity $C_{5\%} / C$ thus 
represents the conditional probability
that, given that a paper is commented, this paper
is within the top $5\%$.
We compare the measured fraction of commented papers within
the top $5\%$ with the distribution of these
values calculated in a statistical model 
where commented and non commented papers
are equally likely to be in the top $5\%$. 
The horizontal lines inside the boxes denote the median value ($50\%$)
of the distribution, boxes delimit the $68.2\%$ confidence intervals
(i.e., one standard deviation), and error bars
denote the $95.4\%$ confidence intervals (i.e., two standard deviations).
}
\end{figure}
\noindent Second, we investigate the presence
of commented papers in the population of the most cited articles.
We compare only
papers published in the same journal, and
assign to each paper a score equal to the fraction of articles
published in the same year that received a smaller number of citations.
If the score of a paper is equal to one 
this means that the paper is the most cited publication
of the year, while if the paper's score 
equals zero this means that the paper 
is among the  least cited articles of the year.
Since
the score is not dependent on the publication
year (i.e., the score is 
a distribution-free indicator), 
we can directly compare the impact of papers 
published in different years,
and thus increase the overall number of comparisons and
obtain a more clear statistical picture.
We then perform, using these scores, a ranking between all papers
independently of their year of publication, and 
select the top $5\%$  of publications (or equivalently those
with score larger than $0.95$) in the ranking.
Notice that this is equivalent to the selection
of the top $5\%$ of the most cited papers in each year. 
We finally count the fraction of commented papers 
in the top  $5\%$, and compare the measured values with those expected by 
chance in the hypothesis that being in the top $5\%$ 
and being a commented paper would be
two independent events~\cite{Radicchi2012}. 
A realization of this statistical 
ensemble is obtained by randomly mixing the scores
only among papers 
published in the same year. 
This procedure ensures that the number of
commented papers per year is constant, but removes 
any eventual dependence between citation
numbers and the fact that papers have or have not 
been commented. By generating 
$10,000$ independent realizations, 
we find that commented papers over-populate the 
top $5\%$ of cited papers in each journal (i.e., measured values
are larger than expected median values), and
more importantly that, in the majority of the
journals, these proportions
are at least two standard deviations larger
than what one would expect by chance (Figure~2).
\\
More in detail, by looking at the
rank probability
density of commented papers (Figure S13), we can
observe a general pattern with the following properties:
(i) At low rank positions, commented papers
are less present than expected. This is not surprising
since low rank positions are mainly occupied by uncited papers
or papers
with few citations, and commented papers
are never uncited because they are cited
at least by the comment and the eventual reply.
(ii) At medium rank positions, the distribution
is compatible with the expected uniform distribution.
(iii) At high rank positions, commented papers
are more present than expected.
This means that
commented papers
over-populate the set of the most
cited publications of each journal, and this observation
is statistically incompatible with the assumption
that commented and non commented papers accumulate
citations with the same rates. 
Notice that this does not mean that a paper, in order
to be highly cited, necessarily needs to be 
a commented paper, but that the conditional probability
 to be a top cited 
paper, given that the paper is a commented article,
is much higher than the respective unconditional probability.
For example in {\it Physical Review D},
the unconditional probability to be in the top $5\%$ is $0.05$,
while  the conditional probability for commented papers
is three times larger and equals $0.15$.


\begin{figure}[!ht]
\begin{center}
\includegraphics[width=0.45\textwidth]{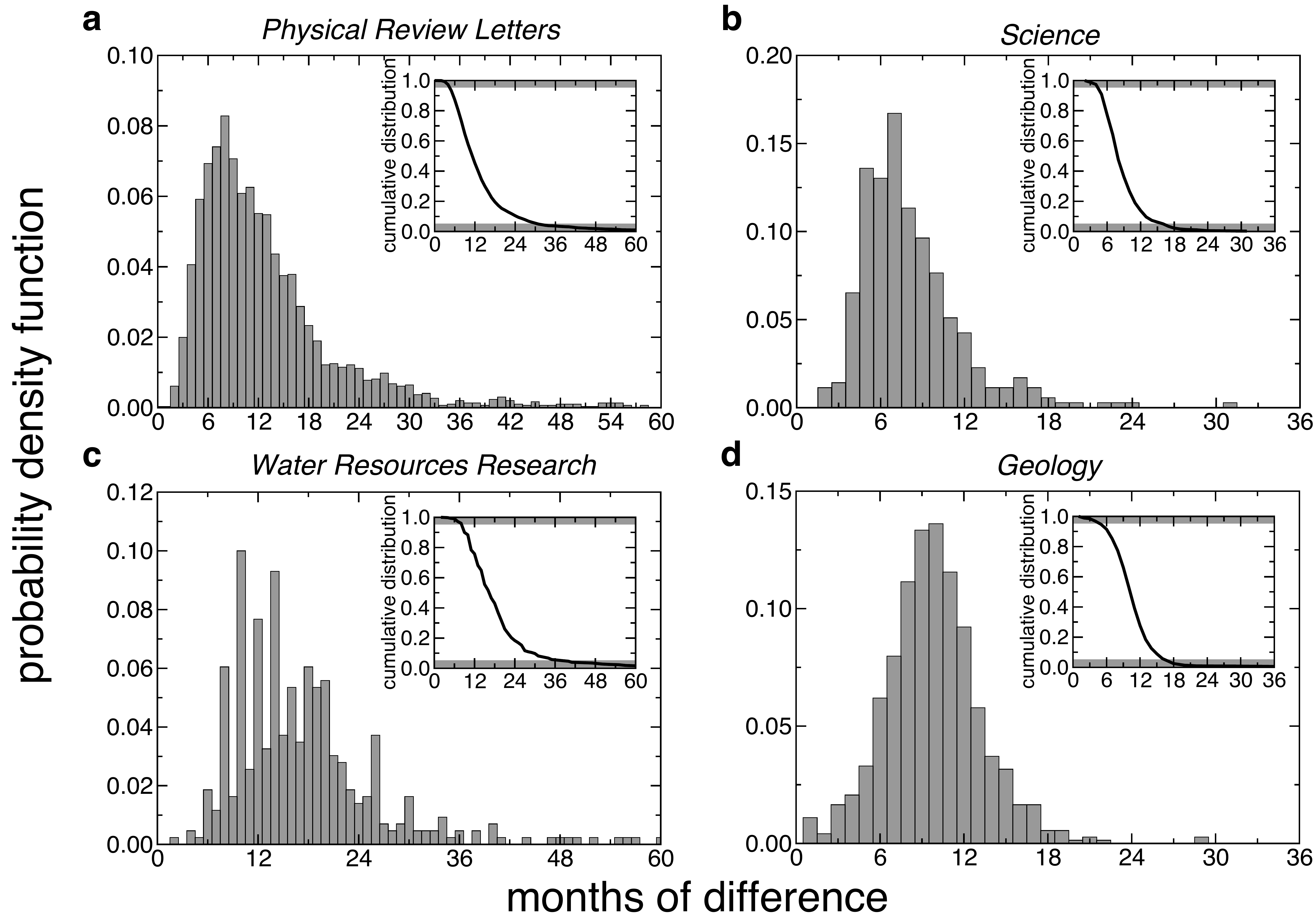}
\end{center}
\caption{
Critical comments follow papers after a short time.
{\bf a.} Probability density function of the difference in publication dates
between comments and commented papers in {\it Physical Review Letters}. Comments
are, on average, published $\tau = 13.5$ months 
later than commented papers (standard deviation $\sigma_\tau = 11.8$). 
In the inset, we report
the cumulative distribution function of the difference
in publication dates between comments and commented papers
in {\it Physical Review Letters}. The $95\%$ of the comments are published
$4-5$ months later than criticized papers, but also in less $30-31$
months after the publication of commented papers. {\bf b.} 
The typical difference
between the publication dates of comments and commented papers in {\it Science}
is $\tau=8.2$ ($\sigma_\tau=3.7$). In the $95\%$ of the cases, 
comments in {\it Science} are published $3-4$ months after the publication of
the commented papers, and before $14-15$ months since the criticized papers
have been published. {\bf c.} In {\it Water Resources Research},  
the time
difference between 
the publication dates of comments and commented papers has an average value
$\tau = 18.3$ ($\sigma_\tau=11.8$). In the $95\%$ of the cases, 
commented papers in {\it Water Resources Research} receive
a comment after $9-10$ months but 
also before $36-40$ months since their publication. {\bf d.} In {\it Geology}, 
papers are commented after an average period $\tau = 10.1$ ($\sigma_\tau=5.4$).
In the $95\%$ of the cases, comments and commented papers 
have publication dates that differ
more than $4-5$ months, but also less than $17-18$ months.
}
\end{figure}

\section{Discussion}

\noindent Contrarily to the popularity of the
wisdom of ``any publicity is good publicity'',
according to which success might follow from negative
criticisms, there are very few empirical validations of 
this belief~\cite{Berger2010}.
In this sense, it is surprising that one example, probably the
most clear empirical observation so far, is indeed offered by science.
It should be noticed that our analysis
does not include all the possible
ways of criticizing previously published
articles. Our approach is in fact limited only to the case
of explicit comments published in the same journal
of publication which published the original, criticized,
article, and therefore neglects possible cases of
``implicit'' criticisms arising from the publication
of other regular articles. This clearly precludes the application
of the same type analysis to disciplines where the publication of 
formal comments is not practiced. 
\\
Also, since comments and commented articles are published at 
short time distance (Figures~3 and S2-10), 
it is difficult to make claims about the causality
effect of a comment on the number
of citations accumulated by a commented paper. 
The time gap is in fact too short to monitor eventual
differences in the trends of accumulation of citations before
and after the publication of a comment, and this
does not allow us to understand
whether highly cited papers attract comments
or instead comments generate citations. 
Our intuitive interpretation is the following.
We think that the potentiality 
of a paper to attract future citations increases
the chances that the paper gets commented. 
Scientists are, in fact, able to early
recognize papers that are predestined to be highly
influential, and thus these papers are more likely to
be deeply examined and eventually commented.
We also think that important scientific 
discoveries often 
originate controversies, and the publication of 
a comment is a way of initiating a dispute. This leads
to boosts of attention and consequently to higher rates of citation
accumulation. 
Independently of the possible interpretation, our
results reveal a coherent pattern 
for all the scientific disciplines
considered here. On the one hand, comments 
can be viewed as negative labels for criticized
papers because comments are published
to criticize or correct other papers. 
On the other hand, comments can be 
also viewed as positive
labels for criticized papers because they 
statistically represent early indicators of
the future impact 
of criticized articles.
\\

\begin{figure}[!ht]
\begin{center}
\includegraphics[width=0.45\textwidth]{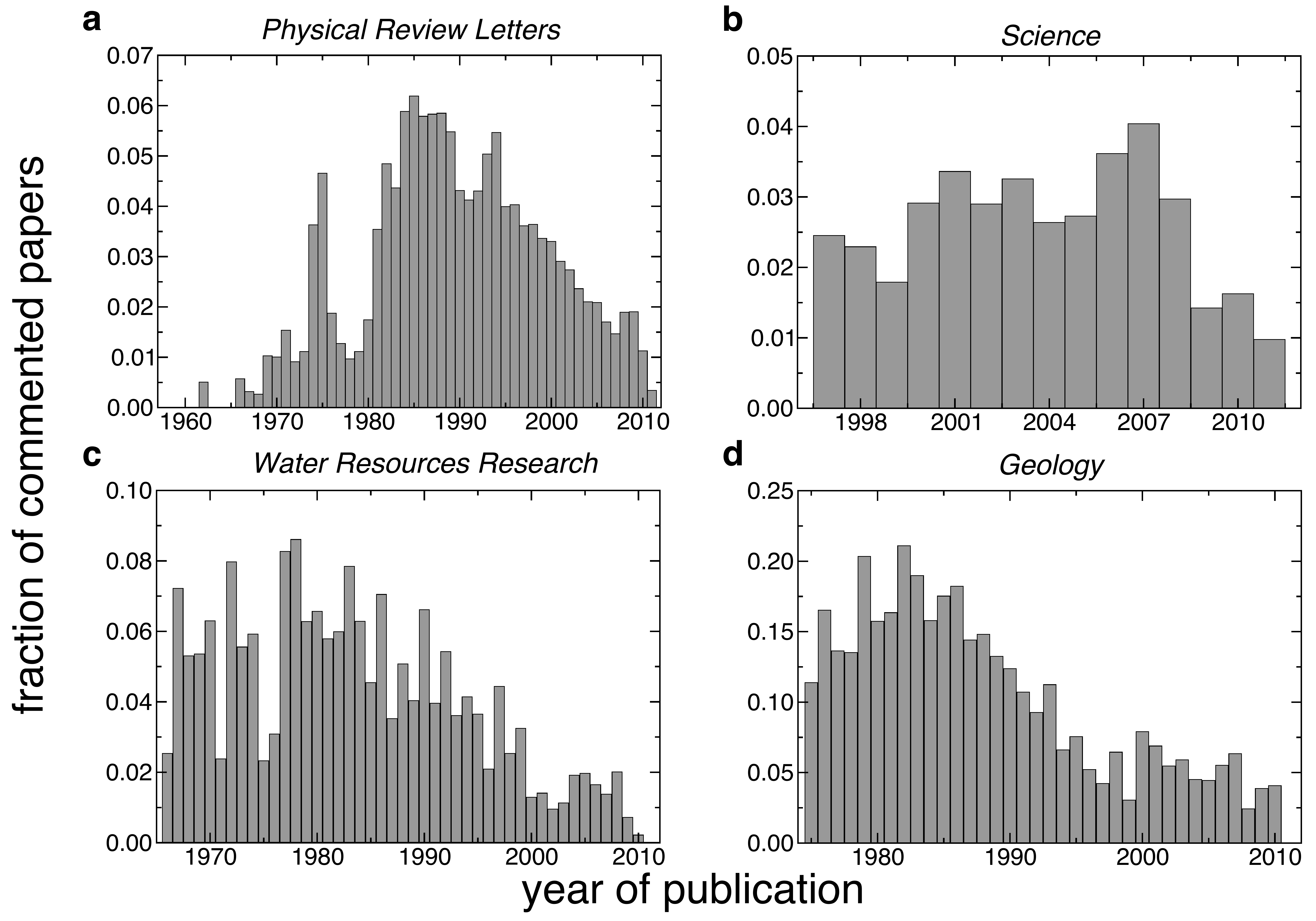}
\end{center}
\caption{
The publication rate of comments has drastically decreased
in recent years. 
Relative number of commented papers
in {\it Physical Review Letters} ({\bf a}), 
{\it Science} ({\bf b}),
{\it Water Resources Research} ({\bf c}) and  {\it Geology} ({\bf d}),
as a function of the year of publication.
}
\end{figure}

\noindent In all journals for which
we could monitor the publication of comments
over sufficiently long time windows, we realized that,
in the latest ten to fifteen years, the proportion
of commented articles has
significantly decreased (Figures~4 and S2-10). 
It seems therefore that, in the process of
creation of scientific knowledge, scientists
are increasingly preferring to avoid
scientific ``fights''.
There could be multiple reasonable explanations
for the observation of such decrement in the
rate of publication of comments, and,
here, we list only few of them. (i) Scientific teams
are increasingly dominating the process
of production of knowledge, and thus,
even if not mutually exclusive, collaboration
is overwhelming antagonism.
(ii) Writing a comment represents an investment of time 
and effort as much as a normal article,
but has low benefit for the academic
curriculum of the authors of the comment
and can be potentially dangerous for the creation
of scientific ``enemies''. (iii) The number of scientific papers
published today is too large. Thus, either scientists have
not enough time to read carefully all published documents
and eventually commenting on them, or the number of
central topics in science that deserve the birth
of a controversy is not growing at the same rate.
(iv) The process of peer-review has become so
precise and efficient that the percentage of published papers
potentially criticizable has decreased.
\\
Clearly, a simple citation analysis, as the one performed here,
cannot provide an exact clue of the reasons for this 
evident reduction. Our study just represents a
starting empirical observation from
which additional questions might arise, and further investigations
are required in order to understand more deeply such
phenomenon. Our analysis, however, provides already a
quantitative recommendation to both scientists
and journal editors to consider the 
fundamental importance of an open scientific 
discussion for the progress of
scientific knowledge. If a paper receives a comment,
this does not necessarily represent a negative event for the journal
which published the paper or for the scientists who wrote
the paper, but could be instead an early indication 
of the importance of the paper itself. As we demonstrated in fact,
if citation numbers truly reflect the scientific impact
of a publication (although this statement is also under
debate~\cite{MacRoberts1996, MacRoberts1989, Bornmann2008, Adler2009}), 
then for a paper it is better 
being commented than not being commented.
Indeed, what Oscar Wilde wrote in the 
Picture of Dorian Gray about gossip seems to be valid also 
in science: ``there is only one thing in the world worse 
than being talked about, and that is not being talked about.''


\newpage

\section*{SUPPLEMENTARY INFORMATION}

\section*{Data collection}

\subsection*{{\it Physical Review}}
For the publication period $1958-2009$, we used the
data set directly provided by the editorial 
office of the American Physical Society 
(APS, {\tt publish.aps.org/datasets}). We identified potential comments
as those publications which satisfy one of the following criteria:
\begin{enumerate}
\item publications whose title starts with the word "Comment" ;
\item publications classified as "comments" in the APS database, but with
title not containing the words "Reply" or "Response".
\end{enumerate}
We then verified for each element in the list of potential comments
whether or not they were effectively comments to previously published papers
in journals of the {\it Physical Review}'s collection. We used two ways
to determine this fact and also to associate real comments to
commented papers:
\begin{enumerate}
\item we parsed the web page of the potential comment, and searched 
for the associated commented article (the criticized article
is referenced in this page as "Original Article"); 
\item when the previous information was 
not present, we read the content of the potential comment and determined
if the document was a real comment with an associated criticized paper.
\end{enumerate}
For years $2010$ and $2011$, we identified potential comments
as all documents with title starting with "Comment on" (the search
was performed in the search engine
of the APS website). We then identified real comments
and associated commented articles in the same way as 
the one described above.
\\
In general, we found that comments were criticizing
papers published in the same journal. We found also
a small percentage of comments criticizing papers
published in other {\it Physical Review} journals. 
Mainly comments to {\it Physical Review Letters} articles
published in the period $1970-1979$ in {\it Physical Review A-D},
and comments to {\it Physical Review A} articles
published  in the period $1993-1994$ in 
{\it Physical Review E} (in $1993$ {\it Physical Review A}  
was split in {\it Physical Review A}  
and in the newly created journal  {\it Physical Review E}).

\subsection*{{\it Science}}
We identified all comments as the elements published by
{\it Science} in the journal section "Technical Comments"
(this list was retrieved at
\\
{\tt www.sciencemag.org/cgi/collection/tech\_comment} 
for publications since $1999$, 
and by reading the content of each issue 
for the previous years). We were
able to consider only publications since 
$1997$ because the electronic journal
archive covers publications from 
$1997$ on (for previous years, publications are provided
only in pdf format). We associated each comment to 
the criticized paper by parsing the web page of the 
on-line version of the comment
and finding the associated document (the criticized paper is listed
after the sentence "The editors suggest the following Related Resources on Science sites"). 
When this information was not present,
we instead read the text of the comment.

\subsection*{{\it Nature}}
For publication year $2004$ and after, we identified all comments as the elements published in
{\it Nature} as "Brief Communications Arising" (such list 
was obtained by parsing the content
of all issues of the journal at {\tt www.nature.com}). 
For years $1999$ to $2003$,
comments have been instead identified (by reading
the text of the publications) among those
papers published in the section "Brief Communications" .
We were able to consider only publications 
since $1999$ because the electronic journal
archive covers publications from $1999$ on (for previous years, publications are provided
only in pdf format). We associated each comment to 
the criticized paper by parsing the web page of the on-line version of the comment
and finding the associated document (the reference to criticized paper
appears at the beginning of the comment after the sentence
"Arising from"). When this information was not present,
we instead read the text of the comment.

\subsection*{{\it New England Journal of Medicine}}
We identified as potential comments all the elements published by
{\it New England Journal of Medicine} in the section 
"Correspondence" (the entire publication list of the correspondence
section can be retrieved with the search engine
offered at {\tt www.nejm.org}).  We were
able to consider only publications since $1990$ because the electronic journal
archive covers publications from $1990$ on (for previous years, publications are provided
only in pdf format). 
We automatically parsed the content of all potential comments, and identified
true comments as those publications in which we identified a string of the
format "(month day issue) reference". The
information contained in this string
was also used to identify the criticized papers.

\subsection*{{\it Journal of Chemical Physics}}
We identified all comments as those publication with title of the format
"Comment on \{title of the commented paper\} 
[{\it J. Chem. Phys.} vol., page (year)]". We limited our attention only
to publications after $1999$ because in previous years
we were not able to detect a regular publication rate of comments.
The reference to the criticized papers was obtained by parsing
the titles of the comments. Data have been collected from the journal 
web site ({\tt jcp.aip.org}).

\subsection*{{\it Geology}}
We identified all comments as those publications with titles
in the format "Comment on \{title of the commented article\}"
or "title of the commented article - Comment". Criticized articles where
automatically detected by matching their titles with those appearing
in the titles of the comments. Data have been collected from the journal web site
({\tt geology.gsapubs.org}).

\subsection*{{\it Environmental Science \& Technology} and {\it Water
Resources Research}}
We identified all comments as those publications with titles
in the format "Comment on \{title of the commented article\}"
or "title of the commented article - Comment". Criticized articles where
automatically detected by matching their titles with those appearing
in the titles of the comments. Data in this case have been retrieved from 
the Web of Science database.


\renewcommand{\thefigure}{S\arabic{figure}}
\setcounter{figure}{0}
\renewcommand{\thetable}{S\arabic{table}}
\setcounter{table}{0}

\begin{table*}
\begin{center}
\begin{tabular}{l r r r r r r}
\hline
Journal & Period of observation & $N$ & $C$ & $P_c$ & $T$ & $r_c$\\
\hline
\hline
{\it Environ. Sci. \& Technol.} & $1981-2007$ & $13,626$ & $206$ & $1.51\%$ & $152$ & $2.3$\\
\hline
{\it Geology} & $1975-2007$ & $7,498$ & $723$ & $9.64\%$ & $9$ & $5.8$\\
\hline
{\it J. Chem. Phys.} & $1999-2007$ & $23,371$ & $111$ & $0.47\%$ & $285$ & $3.7$\\
\hline
{\it Nature} & $1999-2007$ & $8,956$ & $182$ & $2.03\%$ & $233$ & $1.1$\\
\hline
{\it New Engl. J. Med.} &  $1990-2007$ & $5,082$ & $2,667$ & $52.48\%$ & $5$ & $1.9$\\
\hline
{\it Phys. Rev. A} & $1970-2007$ & $47,195$ & $491$ & $1.04\%$ & $96$ & $5.0$\\
\hline
{\it Phys. Rev. B} & $1970-2007$ & $123,242$ & $748$ & $0.61\%$ & $132$ & $6.2$\\
\hline
{\it Phys. Rev. C} & $1970-2007$ & $27,203$ & $248$ & $0.91\%$ & $176$ & $3.1$\\
\hline
{\it Phys. Rev. D} & $1970-2007$ & $48,816$ & $370$ & $0.76\%$ & $15$ & $44.0$\\
\hline
{\it Phys. Rev. E} & $1993-2007$ & $30,413$ & $209$ & $0.69\%$ & $195$ & $3.7$\\
\hline
{\it Phys. Rev. Lett.} & $1970-2007$ &  $74,157$ & $2,475$ & $3.34\%$ & $16$ & $9.4$\\
\hline
{\it Science} &  $1997-2007$ & $9,762$ & $281$ & $2.88\%$ & $74$ & $2.3$\\
\hline
{\it Water Resour. Res.} & $1966-2007$ & $9,487$ & $368$ & $3.88\%$ & $13$ & $9.9$\\
\hline
\hline
\end{tabular}
\end{center}
\caption{Summary table.
For each journal, we report from left to right:
the name of the journal, our observation window, 
the number $N$ of papers published,
the number $C$ of papers that received at least a comment, the percentage $P_c$
of commented papers, the rank $T$ of the fifth most cited
commented paper, and the number of commented
papers in the top $T$ most cited papers of the journal
divided by the expected number of such commented papers
in the hypothetical situation in which commented and commented
papers would not differ in the ability to accumulate
citations. The latter number is simply calculated as
$r_c = \left(5\, N \right) / \left(T\, C \right) $
since $\langle t \rangle = \left(T\, C \right) / N$ is the
expected number of commented articles in the top $T$
most cited papers of the journal.
}
\end{table*}


\begin{table*}
\small
\begin{center}
\begin{tabular}{r r l l}
{\bf rank} & {\bf citations} & {\bf reference} & {\bf commented in}\\
\hline \hline
$1$ & $21,224$ & {\it Phys. Rev. Lett.} {\bf 77}, 3865 (1996)& {\bf 80}, 890 (1998) \\
$3$ & $6,968$ & {\it Phys. Rev. Lett.} {\bf 56}, 930 (1986)& {\bf 57}, 3235 (1986)\\
$5$ & $5,540$ & {\it Phys. Rev. Lett.} {\bf 55}, 2471 (1985)&  {\bf 56}, 2656 (1986)\\
$12$ & $3,946$ & {\it Phys. Rev. Lett.} {\bf 85}, 3966 (2000)&  {\bf 87}, 249701 (2001) \&  {\bf 87}, 249703 (2001)\\
$16$ & $3,331$ & {\it Phys. Rev. Lett.} {\bf 59}, 381 (1987)& {\bf 62}, 110 (1989) \\
\hline
$2$ & $6,188$ & {\it Science} {\bf 286}, 509 (1999)& {\bf 287}, 2115 (2000)\\
$11$ & $3,644$ &  {\it Science} {\bf 280},  69 (1998)& {\bf 281}, 883 (1998) \\
$53$ & $2,085$ & {\it Science} {\bf 290},  2319 (2000)& {\bf 295}, 7 (2002) \\
$60$ & $1,956$ & {\it Science} {\bf 282},  1318 (1998)& {\bf 287}, 1363 (2000)\\
$74$ & $1,690$ &  {\it Science} {\bf 299},  1719 (2003)& {\bf 307}, 1203 (2005)\\
\hline
$2$ & $1,730$ & {\it Water Resour. Res.} {\bf 16}, 574 (1980) & {\bf 17}, 768 (1981)\\
$4$ & $1,032$  & {\it Water Resour. Res.} {\bf 14}, 601 (1978) & {\bf 15}, 989 (1979)\\
$5$ & $986$ & {\it Water Resour. Res.} {\bf 19}, 161 (1983) & {\bf 19}, 1641 (1983)\\
$8$ & $645$  & {\it Water Resour. Res.} {\bf 22}, 2069 (1986) & 
\parbox[t]{5.5cm}{{\bf 24}, 315 (1988) \& {\bf 24}, 892 (1988)\\ \& {\bf 24}, 1209 (1988)}\\
$13$ & $525$ & {\it Water Resour. Res.} {\bf 11}, 725 (1975) & {\bf 13}, 477 (1977)\\
\hline
$4$ & $521$ & {\it Geology} {\bf 19}, 598 (1991) & {\bf 20}, 191 (1992)\\
$5$ & $520$ & {\it Geology} {\bf 19}, 547 (1991) & {\bf 20}, 475 (1992)\\
$6$ & $517$ & {\it Geology} {\bf 10}, 70 (1982) & {\bf 11}, 428 (1993)\\
$7$ & $493$ & {\it Geology} {\bf 19}, 425 (1991) & {\bf 20}, 87 (1993)\\
$9$ & $484$ & {\it Geology} {\bf 14}, 99 (1986) & {\bf 14}, 1042 (1986)\\
\end{tabular}
\end{center}

\caption{Top cited papers that received a comment.
We report here the five top cited papers, among those that received at least
one comment, published in {\it Physical Review Letters} in the period
$1970-2007$, in {\it Science} between $1997$ and $2007$,
in {\it Water Resources Research} in the period $1966-2007$ and 
in {\it Geology} between $1975$ and $2007$.
For each paper, we report the following information: 
absolute rank based on the raw number of citations accumulated (the ranking includes
all publications, either commented and not commented), 
total number of citations accumulated, reference of the commented publication,
reference of the comment(s).}


\end{table*}



\begin{table*}
\small
\begin{center}
\begin{tabular}{r r l l}
{\bf rank} & {\bf citations} & {\bf reference} & {\bf commented in}\\
\hline \hline
$32$ & $743$ & {\it Phys. Rev. A} {\bf 29}, 2765 (1984)& {\bf 32}, 3135 (1985) \\
$49$ & $609$ & {\it Phys. Rev. A} {\bf 39}, 1665 (1989)& {\bf 43}, 2576 (1991)\\
$65$ & $533$ & {\it Phys. Rev. A} {\bf 16}, 531 (1977)&  {\bf 26}, 3008 (1982)\\
$84$ & $479$ & {\it Phys. Rev. A} {\bf 41}, 2295 (1990)&  {\bf 43}, 5165 (1991)\\
$96$ & $447$ & {\it Phys. Rev. A} {\bf 26}, 2028 (1982)& {\bf 36}, 5463 (1987) \\
\hline
$3$ & $11,090$ & {\it Phys. Rev. B} {\bf 13}, 5188 (1976)& {\bf 18}, 5897 (1978) \\
$16$ & $2,557$ & {\it Phys. Rev. B} {\bf 26}, 4199 (1982)& {\bf 37}, 4795 (1988)\\
$31$ & $1,693$ & {\it Phys. Rev. B} {\bf 12}, 2455 (1975)&  {\bf 16}, 4719 (1977)\\
$95$ & $917$ & {\it Phys. Rev. B} {\bf 4}, 3184 (1971)&  {\bf 6}, 311 (1972) \& {\bf 6}, 3546 (1972)\\
$132$ & $813$ & {\it Phys. Rev. B} {\bf 20}, 624 (1979)& {\bf 22}, 1095 (1980) \\
\hline
$24$ & $420$ & {\it Phys. Rev. C} {\bf 13}, 1226 (1976)& {\bf 16}, 885 (1977) \\
$65$ & $277$ & {\it Phys. Rev. C} {\bf 35}, 1678 (1987)& {\bf 37}, 892 (1988)\\
$117$ & $207$ & {\it Phys. Rev. C} {\bf 1}, 769 (1970)&  {\bf 5}, 1135 (1972)\\
$141$ & $194$ & {\it Phys. Rev. C} {\bf 15}, 1359 (1977)&  {\bf 18}, 573 (1978)\\
$176$ & $177$ & {\it Phys. Rev. C} {\bf 9}, 1018 (1974)& {\bf 12}, 686 (1975) \\
\hline
$2$ & $2,916$ & {\it Phys. Rev. D} {\bf 10}, 2445 (1974)& {\bf 12}, 3343 (1975) \\
$3$ & $2,610$ & {\it Phys. Rev. D} {\bf 2}, 1285 (1970)& {\bf 3}, 1043 (1971) \& {\bf 4}, 1918 (1971)\\
$4$ & $2,573$ & {\it Phys. Rev. D} {\bf 7}, 1888 (1973)& 
\parbox[t]{5.5cm}{{\bf 9}, 1129 (1974) \& {\bf 11}, 2332 (1975)\\ \& {\bf 11}, 3040 (1975)}  \\
$7$ & $1,984$ & {\it Phys. Rev. D} {\bf 9}, 3471 (1974)&  {\bf 12}, 923 (1975) \& {\bf 12}, 4006 (1975)\\
$15$ & $1,521$ & {\it Phys. Rev. D} {\bf 14}, 870 (1976)& {\bf 18}, 609 (1978) \\
\hline
$32$ & $354$ & {\it Phys. Rev. E} {\bf 68}, 011306 (2003)& {\bf 70}, 043301 (2004) \\
$72$ & $274$ & {\it Phys. Rev. E} {\bf 48}, R29 (1993)& {\bf 51}, 2669 (1995)\\
$131$ & $206$ & {\it Phys. Rev. E} {\bf 54}, 3853 (1996)&  {\bf 60}, 1099 (1999)\\
$150$ & $196$ & {\it Phys. Rev. E} {\bf 65}, 041903 (1974)&  {\bf 67}, 063901 (2003) \& {\bf 72}, 063901 (2005)\\
$195$ & $165$ & {\it Phys. Rev. E} {\bf 50}, 2064 (1994)& {\bf 53}, 2992 (1996) \\
\end{tabular}
\end{center}
\caption{Top cited papers that received a comment.
We report here the five top cited papers, among those that received at least
one comment, published in {\it Physical Review A-D} in the period
$1970-2007$, and in {\it Physical Review E} between $1993$ and $2007$.
For each paper, we report the following information: 
absolute rank based on the raw number of citations accumulated (the ranking includes
all publications, either commented and not commented), 
total number of citations accumulated, reference of the commented publication,
reference of the comment(s).
}
\end{table*}



\begin{table*}
\small
\begin{center}
\begin{tabular}{r r l l}
{\bf rank} & {\bf citations} & {\bf reference} & {\bf commented in}\\
\hline \hline
$1$ & $2,205$ & {\it Environ. Sci. Technol.} {\bf 36}, 1202 (2002) & 
\parbox[t]{5.5cm}{{\bf 36}, 4003 (2002) \& {\bf 36}, 4005 (2002)\\ \& {\bf 37}, 1052 (2003)}\\
$2$ & $950$ & {\it Environ. Sci. Technol.} {\bf 32}, 1549 (1998) & {\bf 33}, 369 (1998)\\  
$74$ & $328$ & {\it Environ. Sci. Technol.} {\bf 30}, 2432 (1996) & {\bf 31}, 1577 (1997)\\
$122$ & $270$ & {\it Environ. Sci. Technol.} {\bf 27}, 961 (1993) & {\bf 28}, 366 (1994) \&  {\bf 28}, 367 (1994)\\
$152$ & $251$ & {\it Environ. Sci. Technol.} {\bf 30}, 881 (1996) & {\bf 30}, 3132 (1996)\\
\hline
$1$ & $8,834$ & {\it New Engl. J. Med.} {\bf 329}, 977 (1993)& 
\parbox[t]{5.5cm}{{\bf 329}, 1661 (1993) \&\\ 3 comm. in {\bf 330}, 641 (1994)} \\
$2$ & $4,866$ & {\it New Engl. J. Med.} {\bf 333}, 1301 (1995)& 3 comm. in {\bf 334}, 1333 (1996)\\
$3$ & $4,827$ & {\it New Engl. J. Med.} {\bf 338}, 853 (1998)& {\bf 339}, 405 (1998)\\
$4$ & $4,676$  & {\it New Engl. J. Med.} {\bf 346}, 393 (2002)&  4 comm. in {\bf 346}, 1829 (2002)\\
$5$ & $4,339$ & {\it New Engl. J. Med.} {\bf 342}, 145 (2000)& 4 comm. in {\bf 343}, 64 (2000)\\
\hline
$24$ & $457$ & {\it J. Chem. Phys.} {\bf 114}, 5149 (2001) & {\bf 116}, 11039 (2002)\\
$198$ & $149$ & {\it J. Chem. Phys.} {\bf 123}, 164110 (2005) & {\bf 124}, 107101 (2006)\\
$218$ & $144$ & {\it J. Chem. Phys.} {\bf 119}, 12784 (2003) & {\bf 121}, 3347 (2004)\\
$250$ & $135$  & {\it J. Chem. Phys.} {\bf 119}, 2376 (2003) & {\bf 120}, 9427 (2004)\\
$285$ & $129$  & {\it J. Chem. Phys.} {\bf 122}, 204302 (2005) & {\bf 125}, 047101 (2006)\\
\hline
$90$ & $1,260$ & {\it Nature} {\bf 401}, 82 (1999)& {\bf 406}, 367 (2000)\\
$92$ & $1,231$ & {\it Nature} {\bf 434}, 214 (2005)& {\bf 437}, E3  (2005) \&  {\bf 437}, E3 (2005)\\
$96$ & $1,211$ & {\it Nature} {\bf 427}, 145 (2004)& 3 comm. in {\bf 430} issue of 1$^{st}$ July, 2004 \\
$186$ & $940$  & {\it Nature} {\bf 425}, 944 (2003)& in {\bf 429} issue of 13$^{th}$ May, 2004  \\
$233$ & $845$  & {\it Nature} {\bf 423}, 280 (2003)& {\bf 434}, E1  (2005)\\
\end{tabular}
\end{center}
\caption{Top cited papers that received a comment.
We report here the five top cited papers, among those that received at least
one comment, published in {\it Environmental Science \& Technology} 
in the period
$1981-2007$, in {\it New England Journal of Medicine} between $1990$ and $2007$,
in {\it Journal of Chemical Physics} in the period $1999-2007$ and 
in {\it Nature} between $1999$ and $2007$.
For each paper, we report the following information: 
absolute rank based on the raw number of citations accumulated (the ranking includes
all publications, either commented and not commented), 
total number of citations accumulated, reference of the commented publication,
reference of the comment(s).
}
\end{table*}


\begin{figure*}
\begin{center}
\includegraphics[width=0.95\textwidth]{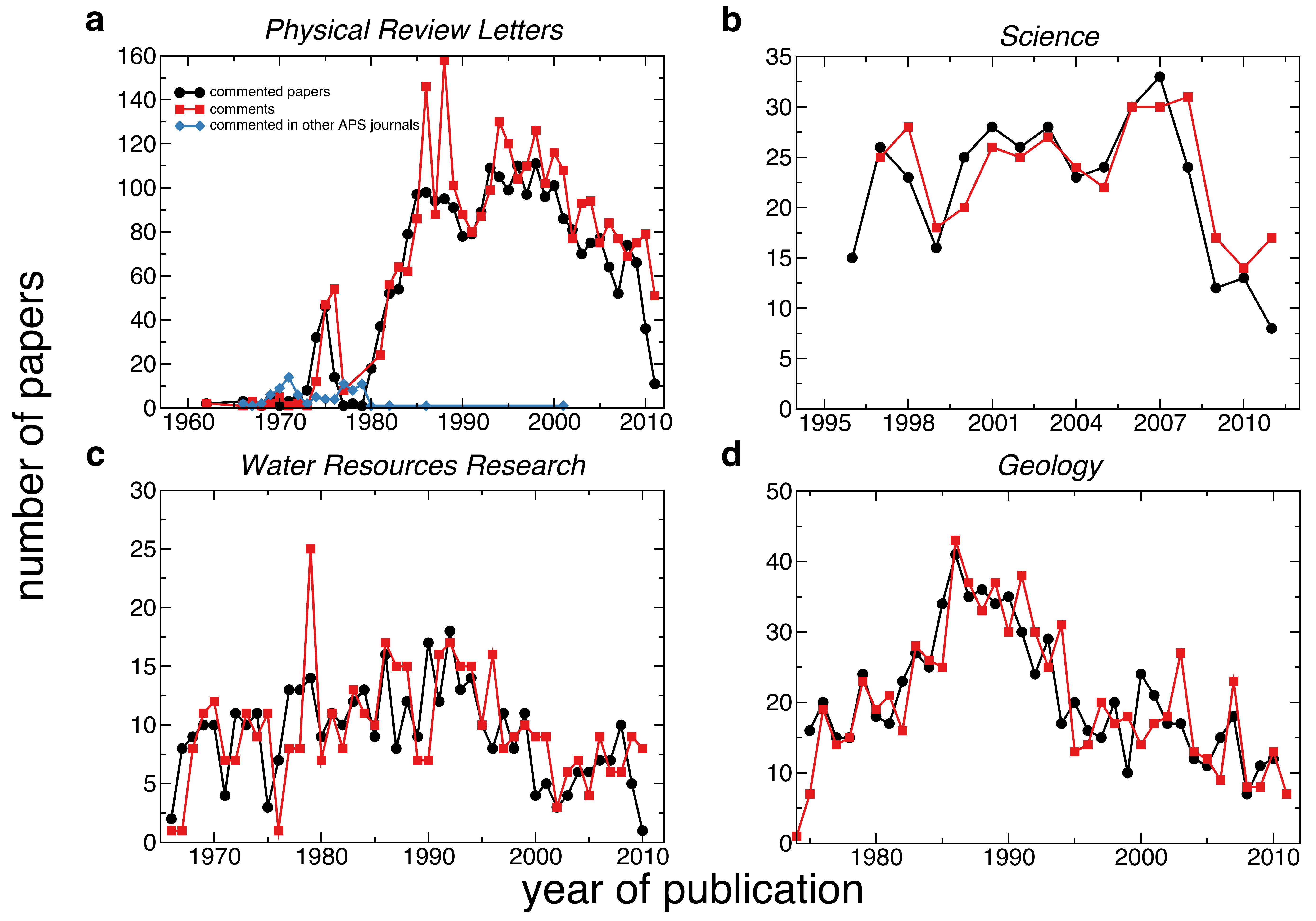}
\end{center}
\caption{Number of commented papers and comments published in each year.
{\bf a.} Number of papers
published in {\it Physical Review Letters} whose comment was published in 
the same journal (black circles) or in other {\it Physical Review} journals (blue diamonds)
as function of the year of publication. Number of comments published
in {\it Physical Review Letters} (red squares) as a function of the year
of publication. {\bf b}, {\bf c} and {\bf d.} Number of commented papers (black circles)
and comments (red squares) published in {\it Science}, 
{\it Water Resources Research} and 
{\it Geology}, respectively.
}
\end{figure*}

\begin{figure*}
\begin{center}
\includegraphics[width=0.95\textwidth]{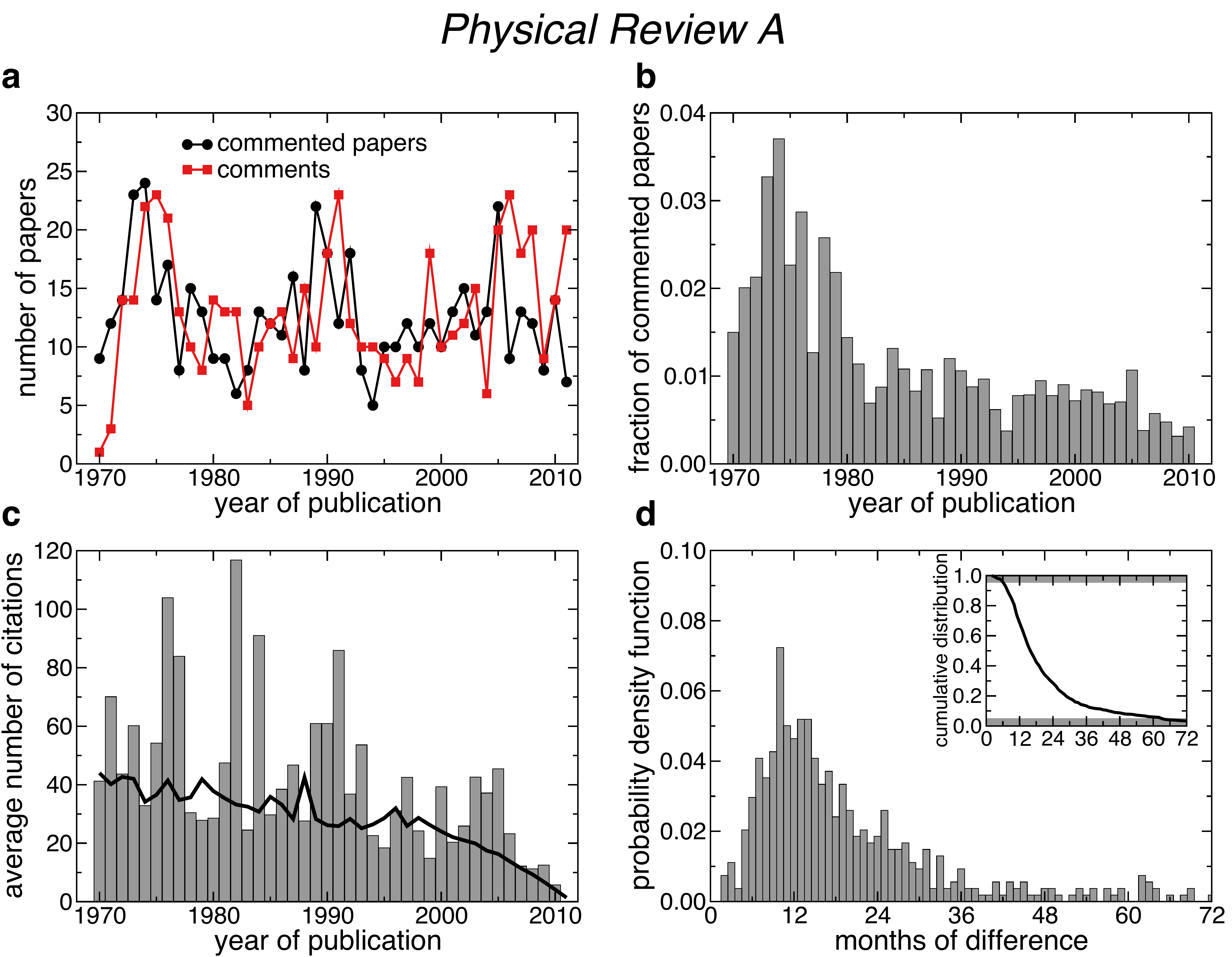}
\end{center}
\caption{Analysis of the publications in {\it Physical Review A}.
{\bf a}. Number of comments (red squares) and commented papers
(black circles) as functions of the year of publication.
{\bf b.} Fraction of commented papers as a function of the year
of publication. {\bf c.} Average number
of citations accumulated by papers published in a given year.
Average citation numbers of commented papers (gray bars)
are compared to those of non commented papers (black line).
{\bf d.} Probability density function (main plot) and
cumulative distribution (inset) of the time difference
between the publication dates of comments and
commented papers. On average, comments are published $\tau = 22.1$
months after commented articles (standard deviation $\sigma_\tau=21.4$). 
}
\end{figure*}

\begin{figure*}
\begin{center}
\includegraphics[width=0.95\textwidth]{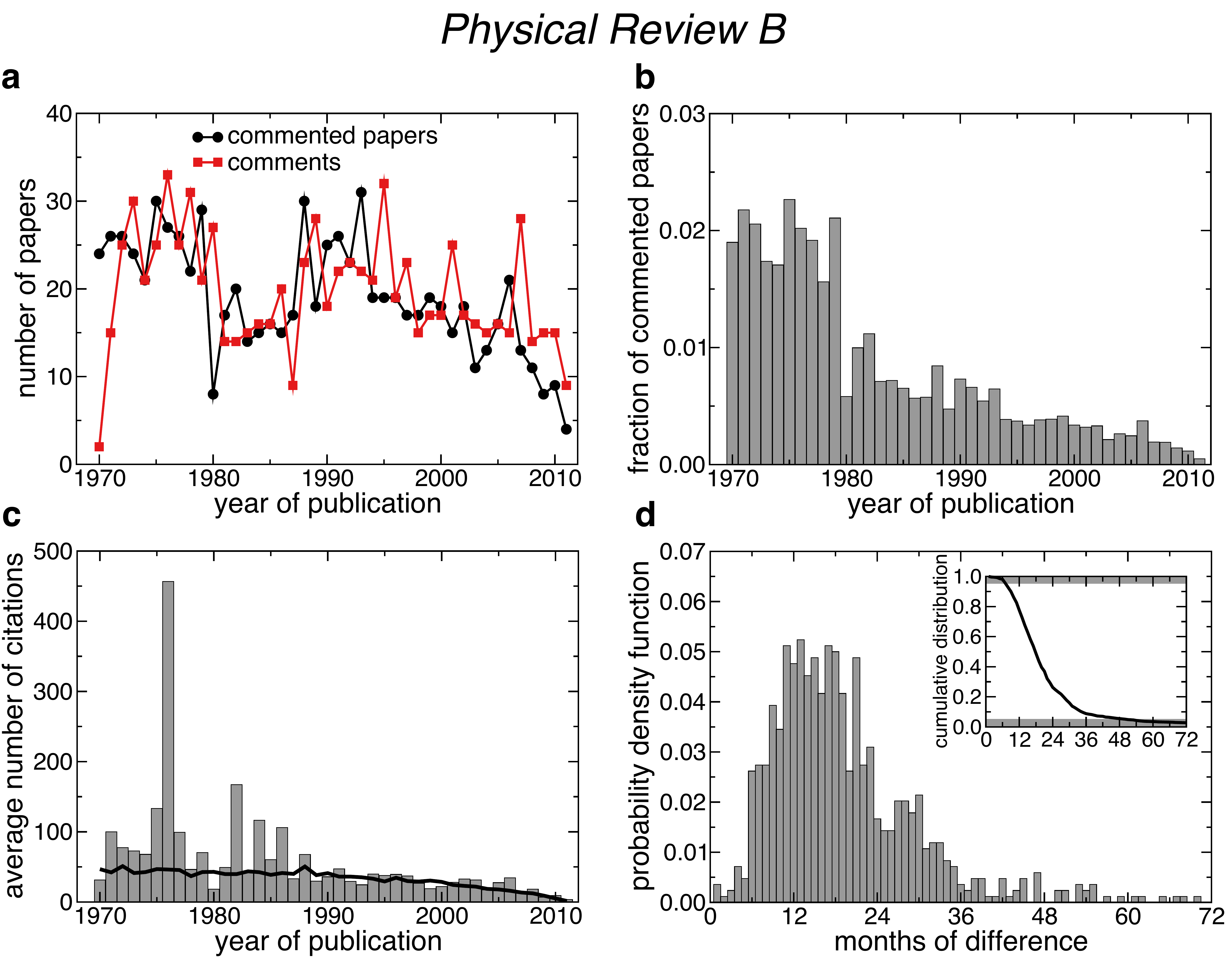}
\end{center}
\caption{Analysis of the publications in {\it Physical Review B}.
{\bf a}. Number of comments (red squares) and commented papers
(black circles) as functions of the year of publication.
{\bf b.} Fraction of commented papers as a function of the year
of publication. {\bf c.} Average number
of citations accumulated by papers published in a given year.
Average citation numbers of commented papers (gray bars)
are compared to those of non commented papers (black line).
{\bf d.} Probability density function (main plot) and
cumulative distribution (inset) of the time difference
between the publication dates of comments and
commented papers. On average, comments are published $\tau = 21.6$
months after commented articles (standard deviation $\sigma_\tau=19.4$). 
}
\end{figure*}

\begin{figure*}
\begin{center}
\includegraphics[width=0.95\textwidth]{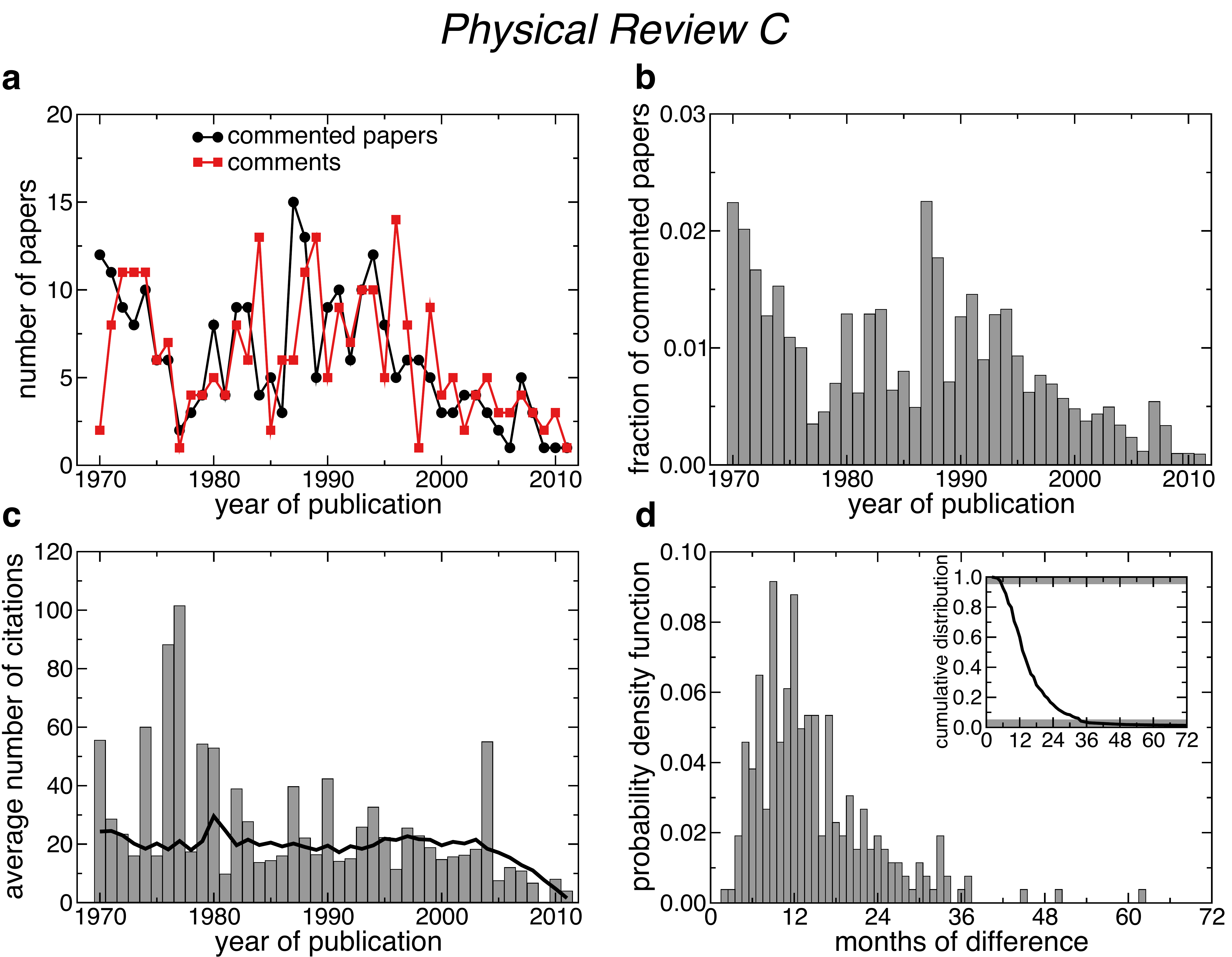}
\end{center}
\caption{Analysis of the publications in {\it Physical Review C}.
{\bf a}. Number of comments (red squares) and commented papers
(black circles) as functions of the year of publication.
{\bf b.} Fraction of commented papers as a function of the year
of publication. {\bf c.} Average number
of citations accumulated by papers published in a given year.
Average citation numbers of commented papers (gray bars)
are compared to those of non commented papers (black line).
{\bf d.} Probability density function (main plot) and
cumulative distribution (inset) of the time difference
between the publication dates of comments and
commented papers. On average, comments are published $\tau = 16.5$
months after commented articles (standard deviation $\sigma_\tau=20.8$). 
}
\end{figure*}

\begin{figure*}
\begin{center}
\includegraphics[width=0.95\textwidth]{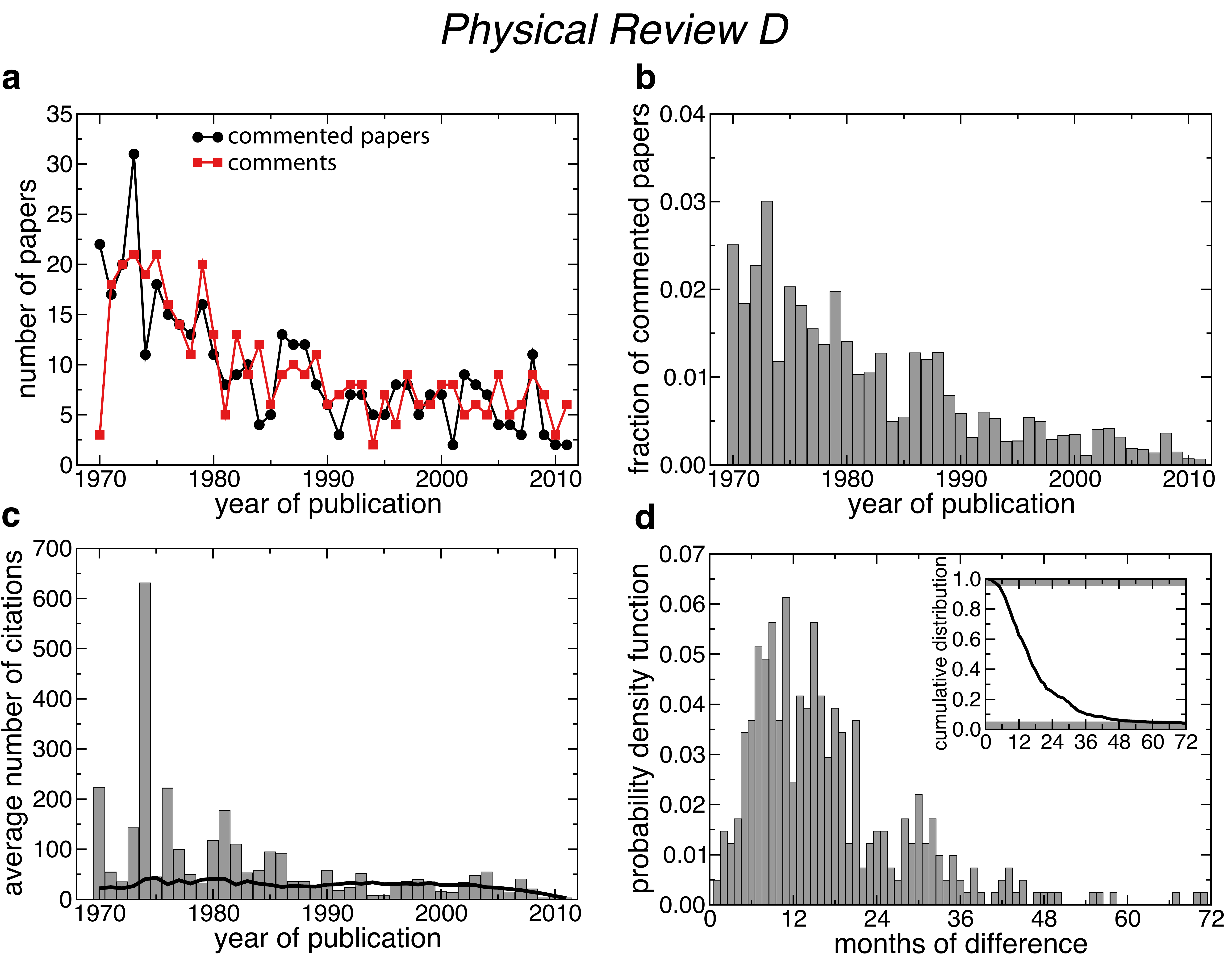}
\end{center}
\caption{Analysis of the publications in {\it Physical Review D}.
{\bf a}. Number of comments (red squares) and commented papers
(black circles) as functions of the year of publication.
{\bf b.} Fraction of commented papers as a function of the year
of publication. {\bf c.} Average number
of citations accumulated by papers published in a given year.
Average citation numbers of commented papers (gray bars)
are compared to those of non commented papers (black line).
{\bf d.} Probability density function (main plot) and
cumulative distribution (inset) of the time difference
between the publication dates of comments and
commented papers. On average, comments are published $\tau = 20.1$
months after commented articles (standard deviation $\sigma_\tau=19.6$). 
}
\end{figure*}

\begin{figure*}
\begin{center}
\includegraphics[width=0.95\textwidth]{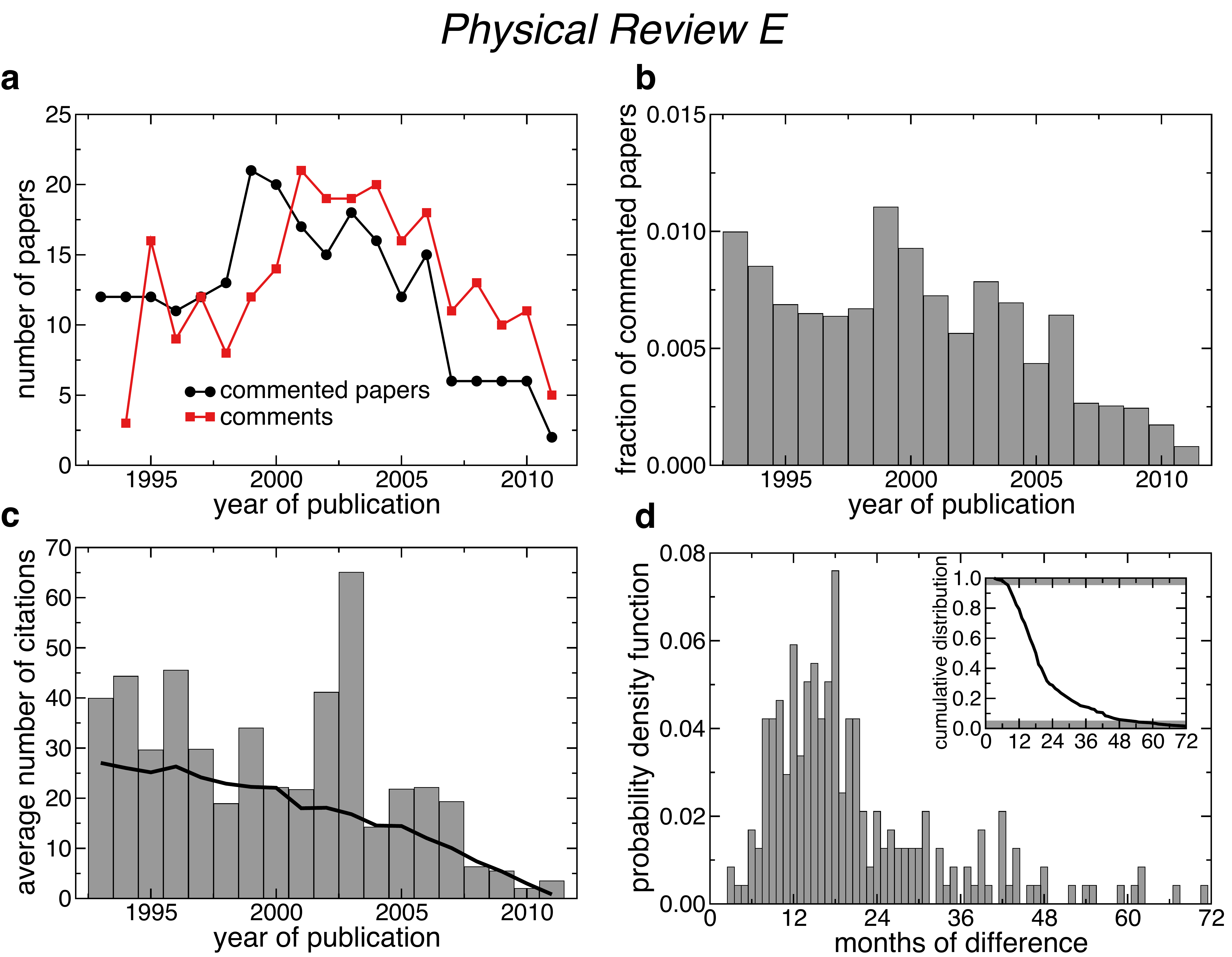}
\end{center}
\caption{Analysis of the publications in {\it Physical Review E}.
{\bf a}. Number of comments (red squares) and commented papers
(black circles) as functions of the year of publication.
{\bf b.} Fraction of commented papers as a function of the year
of publication. {\bf c.} Average number
of citations accumulated by papers published in a given year.
Average citation numbers of commented papers (gray bars)
are compared to those of non commented papers (black line).
{\bf d.} Probability density function (main plot) and
cumulative distribution (inset) of the time difference
between the publication dates of comments and
commented papers. On average, comments are published $\tau = 21.8$
months after commented articles (standard deviation $\sigma_\tau=16.4$). 
}
\end{figure*}

\begin{figure*}
\begin{center}
\includegraphics[width=0.95\textwidth]{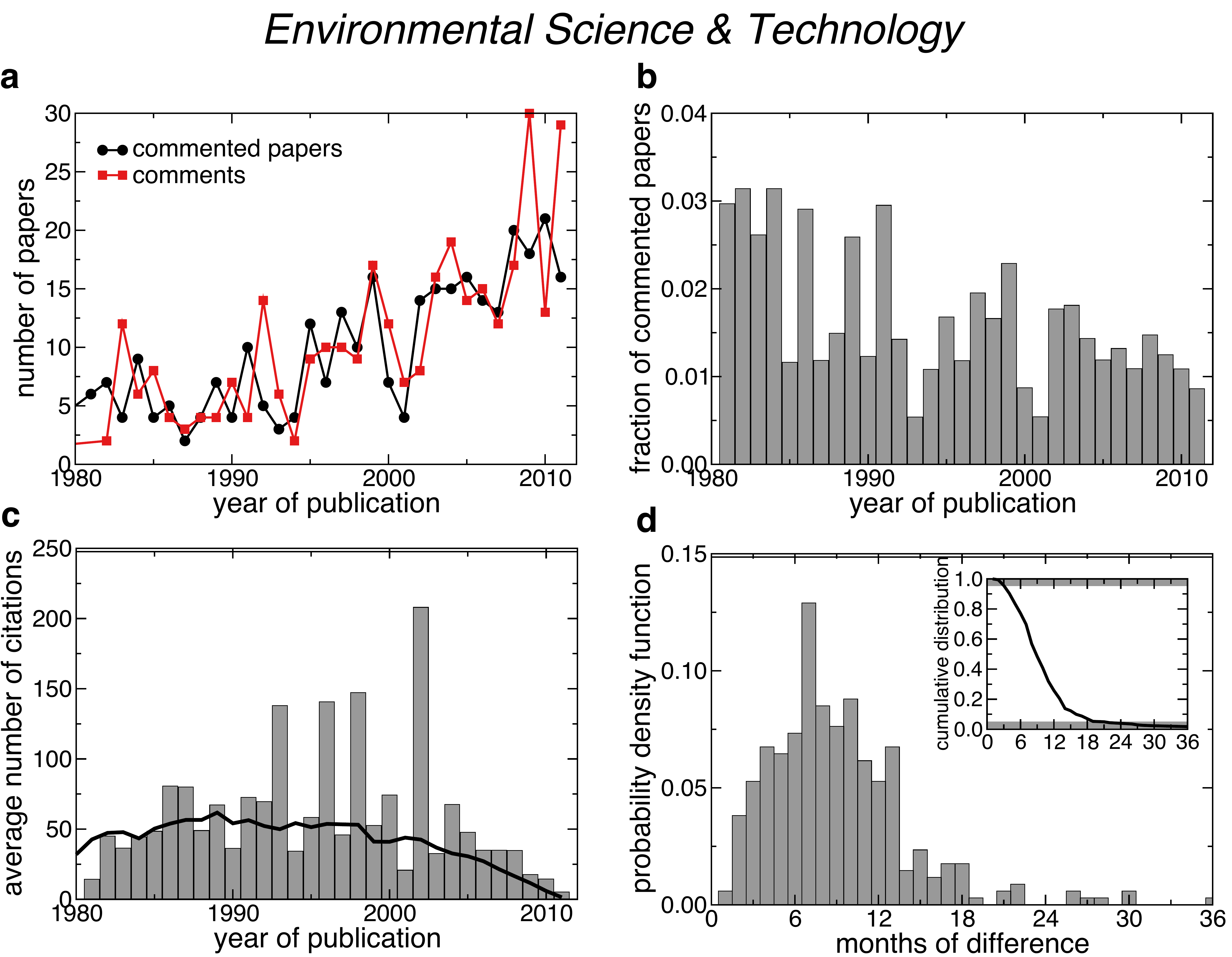}
\end{center}
\caption{Analysis of the publications in {\it Environmental Science \& Technology}.
{\bf a}. Number of comments (red squares) and commented papers
(black circles) as functions of the year of publication.
{\bf b.} Fraction of commented papers as a function of the year
of publication. {\bf c.} Average number
of citations accumulated by papers published in a given year.
Average citation numbers of commented papers (gray bars)
are compared to those of non commented papers (black line).
{\bf d.} Probability density function (main plot) and
cumulative distribution (inset) of the time difference
between the publication dates of comments and
commented papers. On average, comments are published $\tau = 9.6$
months after commented articles (standard deviation $\sigma_\tau=6.5$). 
}
\end{figure*}

\begin{figure*}
\begin{center}
\includegraphics[width=0.95\textwidth]{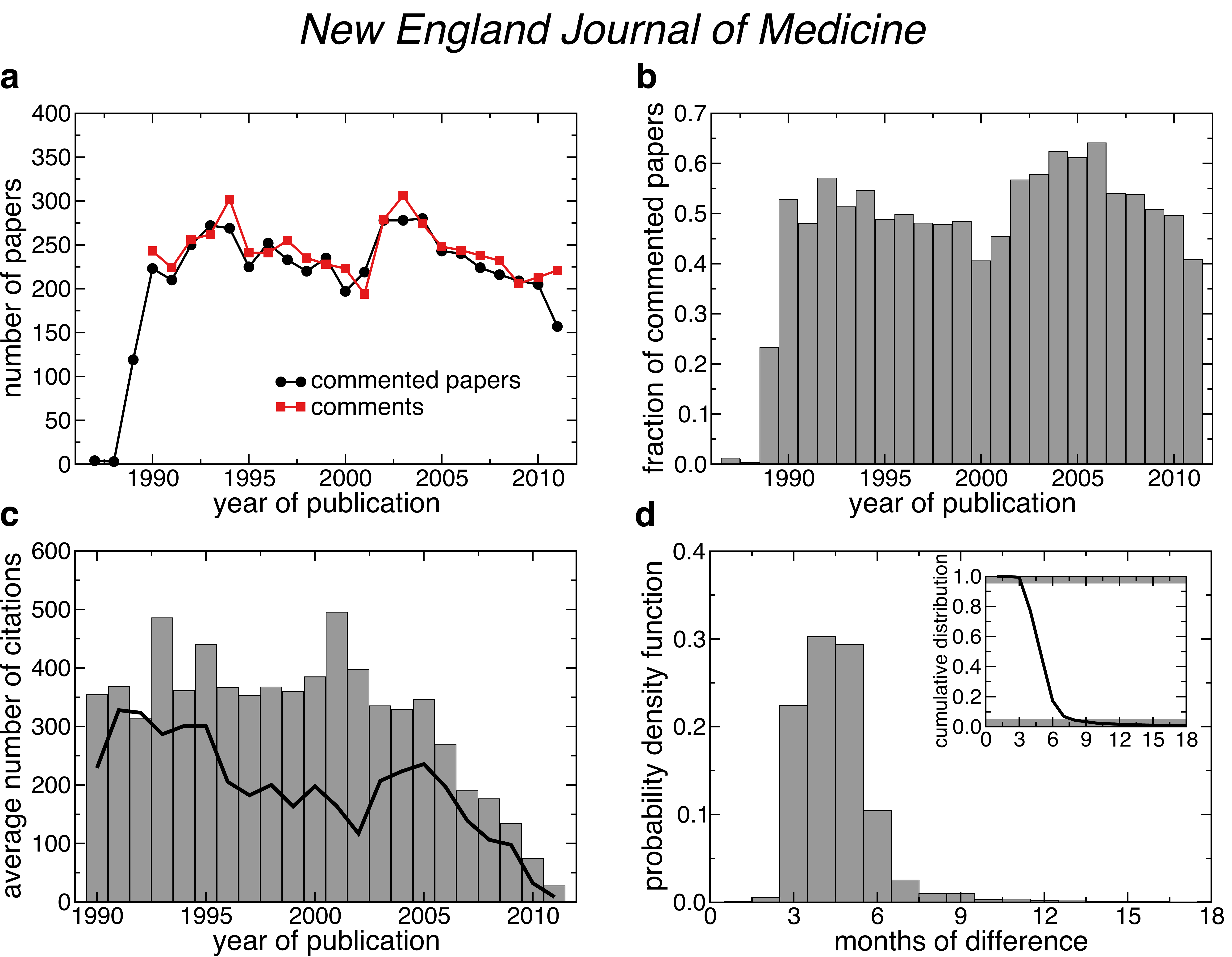}
\end{center}
\caption{Analysis of the publications in {\it New England Journal of Medicine}.
{\bf a}. Number of comments (red squares) and commented papers
(black circles) as functions of the year of publication.
{\bf b.} Fraction of commented papers as a function of the year
of publication. {\bf c.} Average number
of citations accumulated by papers published in a given year.
Average citation numbers of commented papers (gray bars)
are compared to those of non commented papers (black line).
{\bf d.} Probability density function (main plot) and
cumulative distribution (inset) of the time difference
between the publication dates of comments and
commented papers. On average, comments are published $\tau = 4.8$
months after commented articles (standard deviation $\sigma_\tau=3.0$).
}
\end{figure*}

\begin{figure*}
\begin{center}
\includegraphics[width=0.95\textwidth]{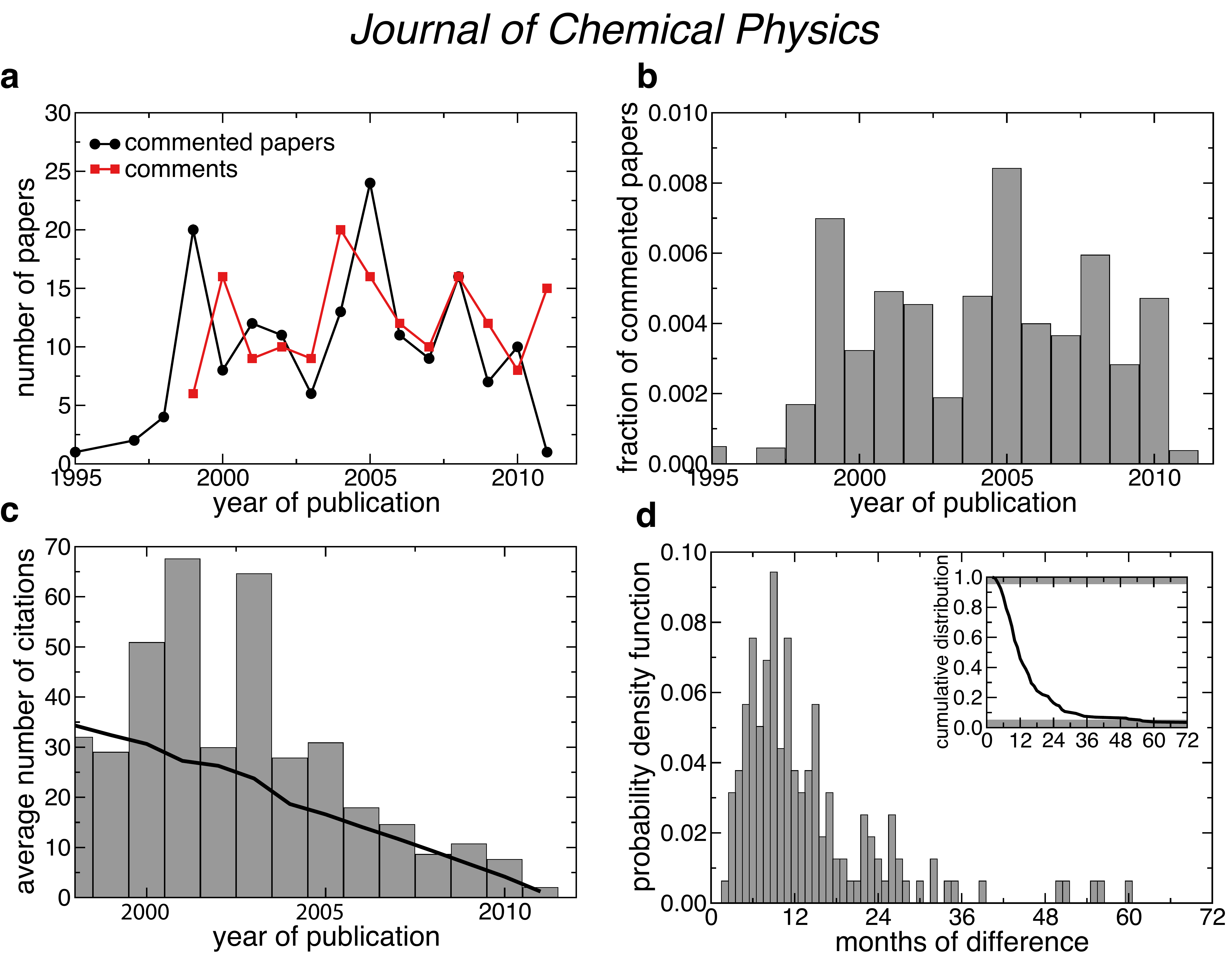}
\end{center}
\caption{Analysis of the publications in {\it Journal of Chemical Physics}.
{\bf a}. Number of comments (red squares) and commented papers
(black circles) as functions of the year of publication.
{\bf b.} Fraction of commented papers as a function of the year
of publication. {\bf c.} Average number
of citations accumulated by papers published in a given year.
Average citation numbers of commented papers (gray bars)
are compared to those of non commented papers (black line).
{\bf d.} Probability density function (main plot) and
cumulative distribution (inset) of the time difference
between the publication dates of comments and
commented papers. On average, comments are published $\tau = 17.6$
months after commented articles (standard deviation $\sigma_\tau=26.3$). 
}
\end{figure*}

\begin{figure*}
\begin{center}
\includegraphics[width=0.95\textwidth]{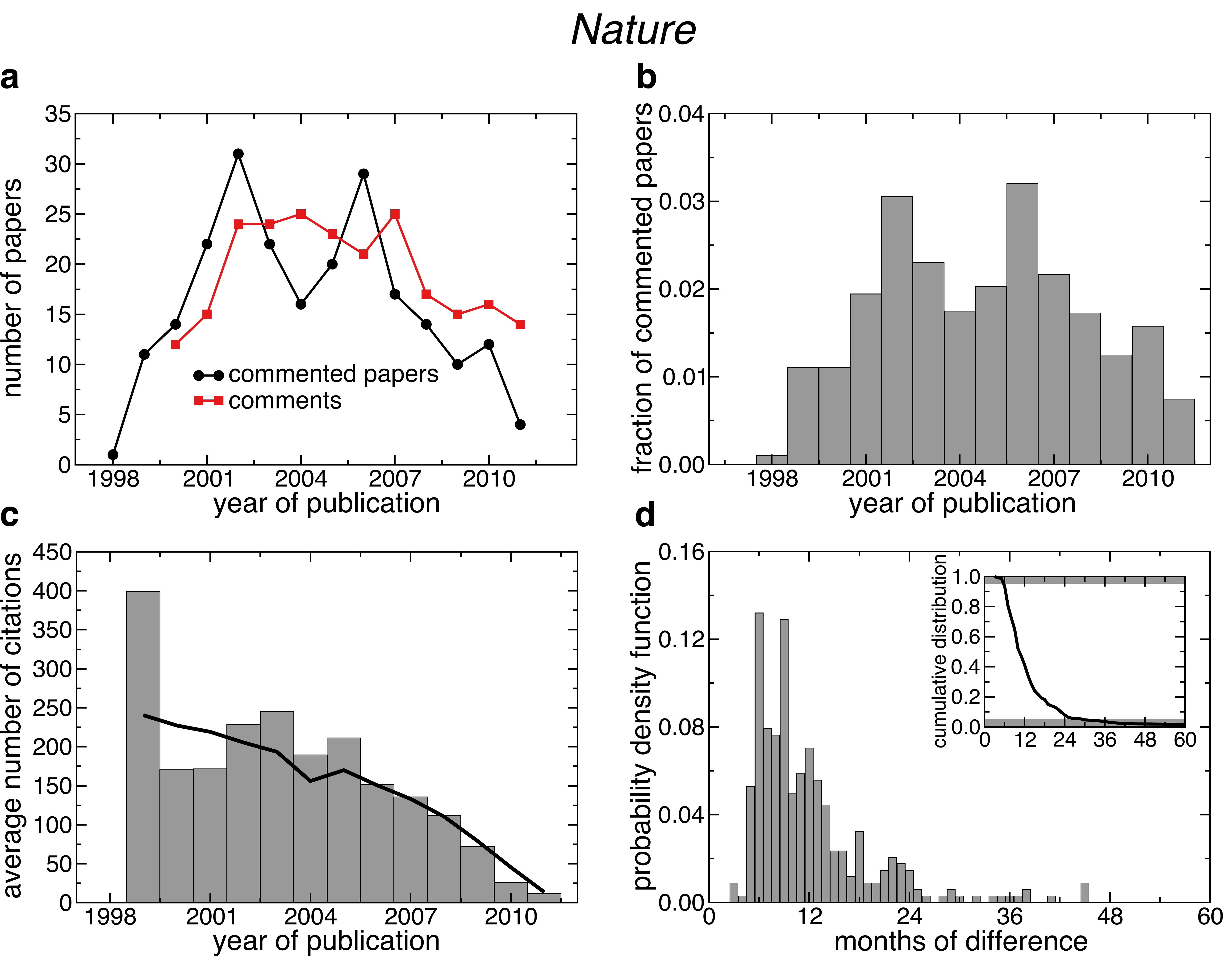}
\end{center}
\caption{Analysis of the publications in {\it Nature}.
{\bf a}. Number of comments (red squares) and commented papers
(black circles) as functions of the year of publication.
{\bf b.} Fraction of commented papers as a function of the year
of publication. {\bf c.} Average number
of citations accumulated by papers published in a given year.
Average citation numbers of commented papers (gray bars)
are compared to those of non commented papers (black line).
{\bf d.} Probability density function (main plot) and
cumulative distribution (inset) of the time difference
between the publication dates of comments and
commented papers. On average, comments are published $\tau = 13.1$
months after commented articles (standard deviation $\sigma_\tau=11.6$). 
}
\end{figure*}

\clearpage

\pagestyle{empty}

\begin{figure*}
\begin{center}
\includegraphics[width=0.95\textwidth]{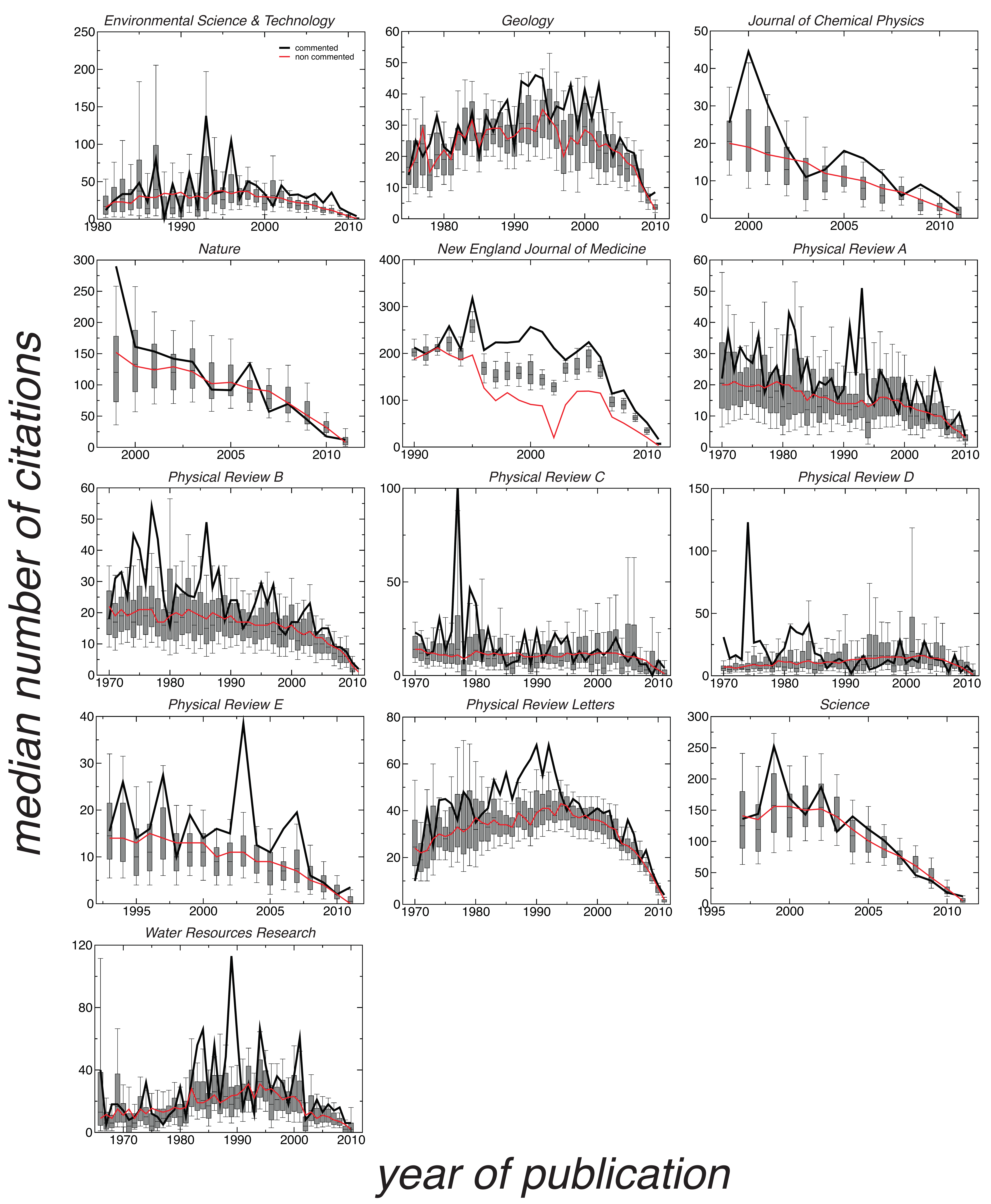}
\end{center}
\caption{Median number of citations as a function of the
publication year of papers. The black line is calculated
for commented papers, while the red line for non commented
ones. The median value of citations accumulated by commented papers
is compared with what expected by chance in
the case in which comments are attributed randomly
to papers. Boxes delimit the $68.2\%$ confidence intervals
(i.e., one standard deviation), while error bars
denote the $95.4\%$ confidence intervals (i.e., two standard deviations).
 }
\end{figure*}

\pagestyle{empty}

\begin{figure*}
\begin{center}
\includegraphics[width=0.95\textwidth]{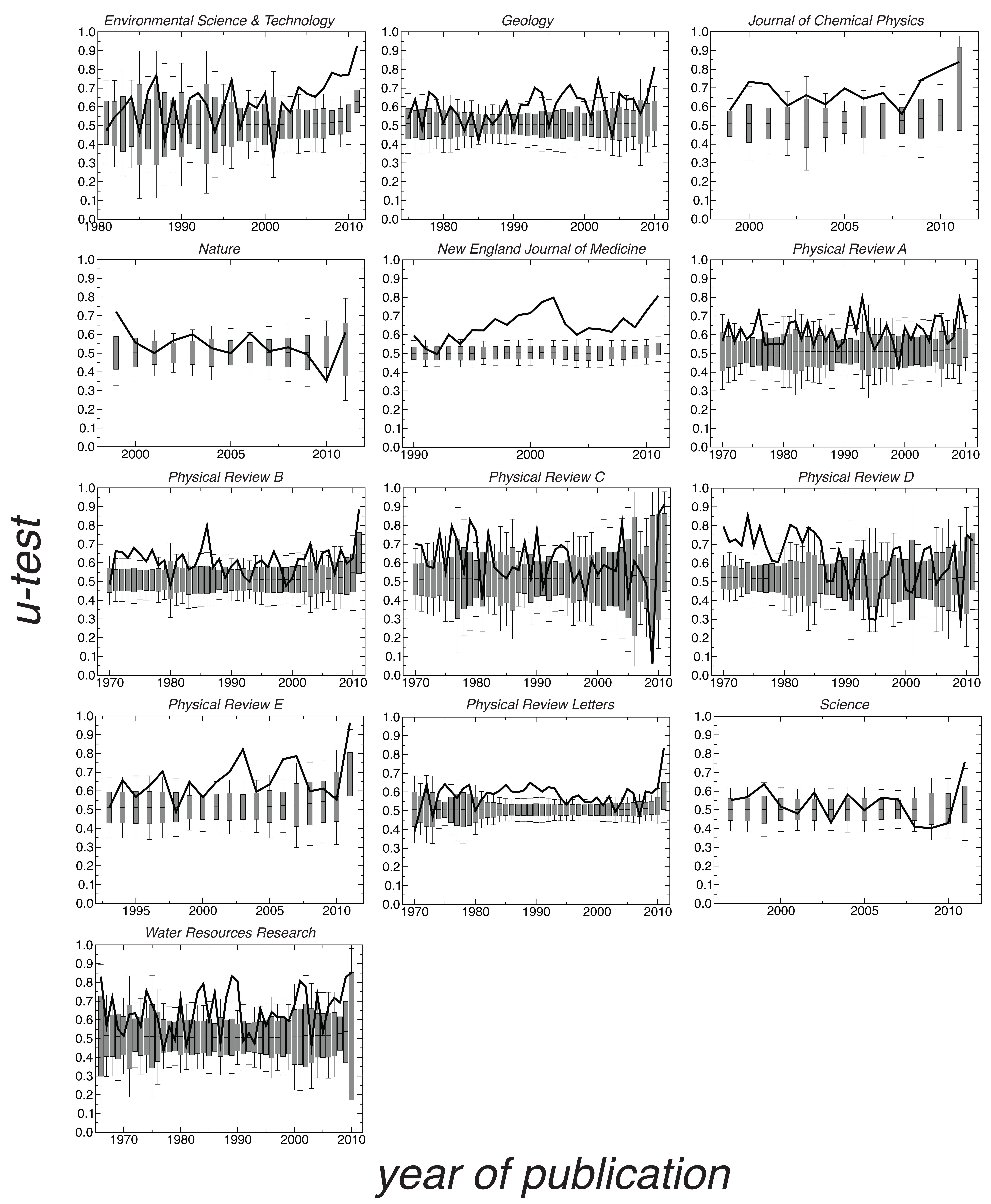}
\end{center}
\caption{Probability
that a commented paper accumulated
at least the same number of citations 
of a non commented paper (u-test). Comparisons
are made only between papers published
in the same year.
Values of the u-test are compared with those expected by chance in
the case in which comments are attributed randomly
to papers. Boxes delimit the $68.2\%$ confidence intervals
(i.e., one standard deviation), while error bars
denote the $95.4\%$ confidence intervals (i.e., two standard deviations).
 }
\end{figure*}

\pagestyle{empty}

\begin{figure*}
\begin{center}
\includegraphics[width=0.95\textwidth]{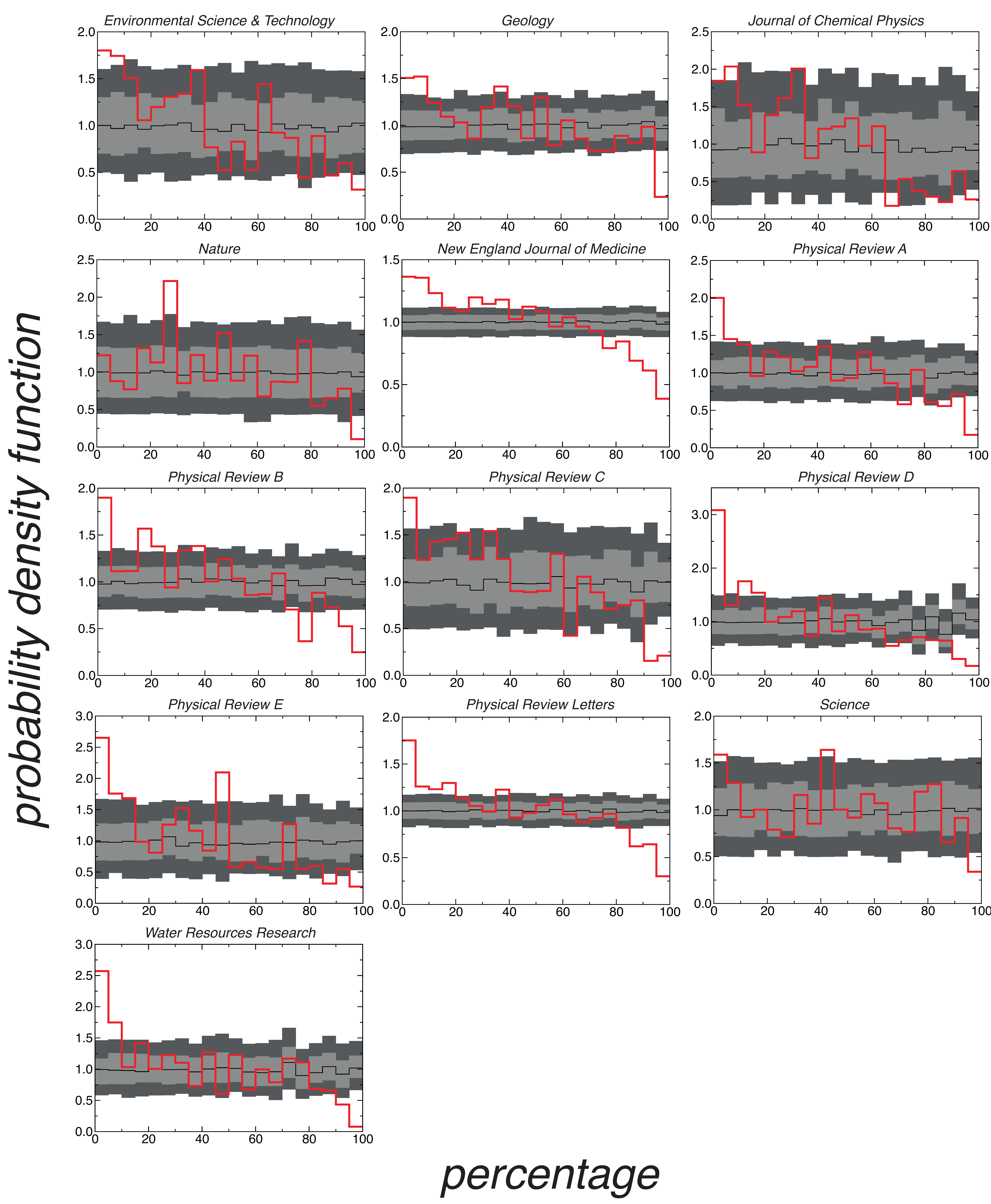}
\end{center}
\caption{Probability
density of commented papers belonging
to a given percentage bin of top cited papers (red curve).
We used bin length equal to $5\%$.
As a term of comparison, we show also the
expected confidence intervals in
the case in which comments are attributed randomly
to papers. The black line show the
median value expected in this case, the light gray region
denotes the $68.2\%$ confidence interval
(i.e., one standard deviation), while the dark gray region
denotes the $95.4\%$ confidence interval (i.e., two standard deviations).
 }
\end{figure*}


\end{document}